%Paper: hep-ph/9308265
%From: trmorris@surya11.cern.ch (Tim Morris)
%Date: Thu, 12 Aug 93 17:41:20 +0200
%Date (revised): Thu, 19 Aug 93 18:26:20 +0200
%Date (revised): Fri, 8 Oct 93 11:16:31 +0100

%   The Exact Renormalisation Group and Approximate Solutions.
%                hep-ph/9308265
%
%                                by  Tim R. Morris
%
%  plain TeX file.  Requires the psfig.tex macro for input of figures.
%  To print the file without the macro. Comment out

\input psfig

%  and comment in

%\def\psfig#1{\vbox{\vskip 3.5in}}

%
%  ( to read text of  paper search ``\preprint'')
%
%
%
%
%\input mac
%
% alias te 'tex exactrg'
% alias po 'dvips -o exactrg.ps exactrg'
% alias pr 'dvips exactrg'
%
%%%%%%% mac  %%%%%%%%%%%%

%%%%%%%%%%%%%%%%  preprint and letter macro  %%%%%%%%%%%%%%%%%

\hsize=6.0truein
\vsize=8.5truein
\voffset=0.25truein
\hoffset=0.1875truein%would be 0.25 except our laser printer is off by 1/16in
\tolerance=1000
\hyphenpenalty=500
\def\monthintext{\ifcase\month\or January\or February\or
   March\or April\or May\or June\or July\or August\or
   September\or October\or November\or December\fi}

%%%%%%%%%%%%%%%%%  Twelve point text font  %%%%%%%%%%%%%%%%%%%

\font\tenrm=cmr10 scaled \magstep1   \font\tenbf=cmbx10 scaled \magstep1
\font\sevenrm=cmr7 scaled \magstep1  
\font\fiverm=cmr5 scaled \magstep1   

\font\teni=cmmi10 scaled \magstep1   \font\tensy=cmsy10 scaled \magstep1
\font\seveni=cmmi7 scaled \magstep1  \font\sevensy=cmsy7 scaled \magstep1
\font\fivei=cmmi5 scaled \magstep1   \font\fivesy=cmsy5 scaled \magstep1

\font\tentt=cmtt10 scaled \magstep1
\font\tenit=cmti10 scaled \magstep1
\font\tensl=cmsl10 scaled \magstep1

\def\twelvepoint{\def\rm{\fam0\tenrm}
   \textfont0=\tenrm \scriptfont0=\sevenrm \scriptscriptfont0=\fiverm
   \textfont1=\teni  \scriptfont1=\seveni  \scriptscriptfont1=\fivei
   \textfont2=\tensy \scriptfont2=\sevensy \scriptscriptfont2=\fivesy
   \textfont\itfam=\tenit \def\it{\fam\itfam\tenit}
   \textfont\ttfam=\tentt \def\tt{\fam\ttfam\tentt}
   \textfont\bffam=\tenbf \def\bf{\fam\bffam\tenbf}
   \textfont\slfam=\tensl \def\sl{\fam\slfam\tensl} \rm
   %Essentially I changed all dimensions to 1.2 times as large as in plain tex
   \hfuzz=1pt\vfuzz=1pt%much more than plain tex's value
   \setbox\strutbox=\hbox{\vrule height 10.2pt depth 4.2pt width 0pt}
   \parindent=24pt\parskip=1.2pt plus 1.2pt
   \topskip=12pt\maxdepth=4.8pt\jot=3.6pt
   \normalbaselineskip=14.4pt\normallineskip=1.2pt
   \normallineskiplimit=0pt\normalbaselines
   \abovedisplayskip=13pt plus 3.6pt minus 5.8pt
   \belowdisplayskip=13pt plus 3.6pt minus 5.8pt
   \abovedisplayshortskip=-1.4pt plus 3.6pt
   \belowdisplayshortskip=13pt plus 3.6pt minus 3.6pt
   %plain tex's value for belowdisplayshortskip looked terrible
   \topskip=12pt \splittopskip=12pt
   \scriptspace=0.6pt\nulldelimiterspace=1.44pt\delimitershortfall=6pt
   \thinmuskip=3.6mu\medmuskip=3.6mu plus 1.2mu minus 1.2mu
   \thickmuskip=4mu plus 2mu minus 1mu%reduced these plain tex values
   \smallskipamount=3.6pt plus 1.2pt minus 1.2pt
   \medskipamount=7.2pt plus 2.4pt minus 2.4pt
   \bigskipamount=14.4pt plus 4.8pt minus 4.8pt}

\twelvepoint

%%%%%%%%%%%%%%%%% Definitions for Preprints %%%%%%%%%%%%%%%%%%

% title page title font

\font\titlerm=cmr10 scaled \magstep3
\font\titlerms=cmr10 scaled \magstep1 %\font\titlermss=cmr8
\font\titlei=cmmi10 scaled \magstep3  %math italic for title
\font\titleis=cmmi10 scaled \magstep1 %\font\titleiss=cmmi8
\font\titlesy=cmsy10 scaled \magstep3 	%math symbols for title
\font\titlesys=cmsy10 scaled \magstep1  %\font\titlesyss=cmsy8
\font\titleit=cmti10 scaled \magstep3	%text italic for title
\skewchar\titlei='177 \skewchar\titleis='177 %\skewchar\titleiss='177
\skewchar\titlesy='60 \skewchar\titlesys='60 %\skewchar\titlesyss='60

\def\titlefont{\def\rm{\fam0\titlerm}% switch to title font
   \textfont0=\titlerm \scriptfont0=\titlerms %\scriptscriptfont0=\titlermss
   \textfont1=\titlei  \scriptfont1=\titleis  %\scriptscriptfont1=\titleiss
   \textfont2=\titlesy \scriptfont2=\titlesys %\scriptscriptfont2=\titlesyss
   \textfont\itfam=\titleit \def\it{\fam\itfam\titleit} \rm}

% title page macros

\def\preprint#1{\baselineskip=19pt plus 0.2pt minus 0.2pt \pageno=0
   \begingroup%use with \draft or \date to end group
   \nopagenumbers\parindent=0pt\baselineskip=14.4pt\rightline{#1}}
\def\title#1{
   \vskip 0.9in plus 0.45in
   \centerline{\titlefont #1}}
\def\secondtitle#1{}%set up this some time
\def\author#1#2#3{\vskip 0.9in plus 0.45in
   \centerline{{\bf #1}\myfoot{#2}{#3}}\vskip 0.12in plus 0.02in}
\def\secondauthor#1#2#3{}%set up this some time
\def\addressline#1{\centerline{#1}}
\def\abstract{\vskip 0.7in plus 0.35in
	\centerline{\bf Abstract}
	\smallskip}
\def\finishtitlepage#1{\vskip 0.8in plus 0.4in
   \leftline{#1}\supereject\endgroup}

\def\date#1{\finishtitlepage{#1}}

\def\nolabels{\def\eqnlabel##1{}\def\eqlabel##1{}\def\figlabel##1{}%
	\def\reflabel##1{}}
\def\writelabels{\def\eqnlabel##1{%
	{\escapechar=` \hfill\rlap{\hskip.11in\string##1}}}%
	\def\eqlabel##1{{\escapechar=` \rlap{\hskip.11in\string##1}}}%
	\def\figlabel##1{\noexpand\llap{\string\string\string##1\hskip.66in}}%
	\def\reflabel##1{\noexpand\llap{\string\string\string##1\hskip.37in}}}
\nolabels

%  tagged section numbers

\global\newcount\secno \global\secno=0
\global\newcount\meqno \global\meqno=1
\global\newcount\subsecno \global\subsecno=0

\font\secfont=cmbx12 scaled\magstep1

\def\section#1{\global\advance\secno by1
   \xdef\secsym{\the\secno.}
   \global\subsecno=0
   \global\meqno=1\bigbreak\medskip
   \noindent{\secfont\the\secno. #1}\par\nobreak\smallskip\nobreak\noindent}
%\xdef\secsym{}

\def\subsection#1{\global\advance\subsecno by1
    %\xdef\secsym{\the\subsecno}
\medskip
\noindent
{\bf\the\secno.\the\subsecno\ #1}
\par\medskip\nobreak\noindent}
%\xdef\secsym{}

\def\newsec#1{\global\advance\secno by1
   \xdef\secsym{\the\secno.}
   \global\meqno=1\bigbreak\medskip
   \noindent{\bf\the\secno. #1}\par\nobreak\smallskip\nobreak\noindent}
\xdef\secsym{}

\def\appendix#1#2{\global\meqno=1\xdef\secsym{\hbox{#1.}}\bigbreak\medskip
\noindent{\bf Appendix #1. #2}\par\nobreak\smallskip\nobreak\noindent}

\def\acknowledgements{\bigbreak\medskip\centerline{\bf
   Acknowledgements}\par\nobreak\smallskip\nobreak\noindent}

%         equations

\def\eqnn#1{\xdef #1{(\secsym\the\meqno)}%
	\global\advance\meqno by1\eqnlabel#1}
\def\eqna#1{\xdef #1##1{\hbox{$(\secsym\the\meqno##1)$}}%
	\global\advance\meqno by1\eqnlabel{#1$\{\}$}}
\def\eqn#1#2{\xdef #1{(\secsym\the\meqno)}\global\advance\meqno by1%
	$$#2\eqno#1\eqlabel#1$$}

%			 footnotes

\def\myfoot#1#2{{\baselineskip=14.4pt plus 0.3pt\footnote{#1}{#2}}}
%sequentially numbered footnotes
\global\newcount\ftno \global\ftno=1
\def\foot#1{{\baselineskip=14.4pt plus 0.3pt\footnote{$^{\the\ftno}$}{#1}}%
	\global\advance\ftno by1}

%         references

\global\newcount\refno \global\refno=1
\newwrite\rfile

\def\ref{[\the\refno]\nref}
\def\nref#1{\xdef#1{[\the\refno]}\ifnum\refno=1\immediate
	\openout\rfile=refs.tmp\fi\global\advance\refno by1\chardef\wfile=\rfile
	\immediate\write\rfile{\noexpand\item{#1\ }\reflabel{#1}\pctsign}\findarg}
%	horrible hack to sidestep tex \write limitation
\def\findarg#1#{\begingroup\obeylines\newlinechar=`\^^M\passarg}
	{\obeylines\gdef\passarg#1{\writeline\relax #1^^M\hbox{}^^M}%
	\gdef\writeline#1^^M{\expandafter\toks0\expandafter{\striprelax #1}%
	\edef\next{\the\toks0}\ifx\next\null\let\next=\endgroup\else\ifx\next\empty%

\else\immediate\write\wfile{\the\toks0}\fi\let\next=\writeline\fi\next\relax}}
	{\catcode`\%=12\xdef\pctsign{%}}
\def\striprelax#1{}

\def\semi{;\hfil\break}
\def\addref#1{\immediate\write\rfile{\noexpand\item{}#1}} %now unnecessary

\def\listrefs{\vfill\eject\immediate\closeout\rfile
   {{\secfont References}}\bigskip{\frenchspacing%
   \catcode`\@=11\escapechar=` %
   \input refs.tmp\vfill\eject}\nonfrenchspacing}

\def\startrefs#1{\immediate\openout\rfile=refs.tmp\refno=#1}

%		and finally, figures:

\global\newcount\figno \global\figno=1
\newwrite\ffile
\def\fig{\the\figno\nfig}
\def\nfig#1{\xdef#1{\the\figno}\ifnum\figno=1\immediate
	\openout\ffile=figs.tmp\fi\global\advance\figno by1\chardef\wfile=\ffile
	\immediate\write\ffile{\medskip\noexpand\item{Fig.\ #1:\ }%
	\figlabel{#1}\pctsign}\findarg}

\def\listfigs{\vfill\eject\immediate\closeout\ffile{\parindent48pt
	\baselineskip16.8pt{{\secfont Figure Captions}}\medskip
	\escapechar=` \input figs.tmp\vfill\eject}}

%%%%%%%%%%%%%%%%%%%%%%%%%%%%%%%%%%%%%%%%%%%%%%%%%%%%%%%%%%%%%%%%%%%%%%%%%%%%%%%
\def\noblackbox{\overfullrule=0pt}
\def\inv{^{\raise.18ex\hbox{${\scriptscriptstyle -}$}\kern-.06em 1}}
\def\dup{^{\vphantom{1}}}
\def\Dsl{\,\raise.18ex\hbox{/}\mkern-16.2mu D} %this one can be subscripted
\def\dsl{\raise.18ex\hbox{/}\kern-.68em\partial}
\def\slash#1{\raise.18ex\hbox{/}\kern-.68em #1}
\def\lspace{}
\def\lbspace{}
\def\boxeqn#1{\vcenter{\vbox{\hrule\hbox{\vrule\kern3.6pt\vbox{\kern3.6pt
	\hbox{${\displaystyle #1}$}\kern3.6pt}\kern3.6pt\vrule}\hrule}}}
\def\mbox#1#2{\vcenter{\hrule \hbox{\vrule height#2.4in
	\kern#1.2in \vrule} \hrule}}  %e.g. \mbox{.1}{.1}
%matters of taste
%\def\tilde{\widetilde}
\def\bar{\overline}
\def\e#1{{\rm e}^{\textstyle#1}}
\def\del{\partial}
\def\curly#1{{\hbox{{$\cal #1$}}}}
\def\curlyD{\hbox{{$\cal D$}}}
\def\curlyL{\hbox{{$\cal L$}}}
\def\vev#1{\langle #1 \rangle}
\def\psibar{\overline\psi}
\def\lform{\hbox{$\sqcup$}\llap{\hbox{$\sqcap$}}}
\def\darr#1{\raise1.8ex\hbox{$\leftrightarrow$}\mkern-19.8mu #1}
\def\half{{\textstyle{1\over2}}} %puts a small half in a displayed eqn
\def\roughly#1{\ \lower1.5ex\hbox{$\sim$}\mkern-22.8mu #1\,}
\def\MSbar{$\bar{{\rm MS}}$}
%%%%%%%%%%%%%%%%%%%%%%%%%%%%%%%%%%%%%%%%%%%%%%%%%%%%%%%%%%%%%%
\hyphenation{di-men-sion di-men-sion-al di-men-sion-al-ly}

\parindent=0pt
\parskip=5pt

%%%%%%%%%%%%%end mac%%%%%%%%%%%%%%%

\preprint{
\vbox{
\rightline{CERN-TH.6977/93}
\vskip2pt\rightline{SHEP 92/93-27}
%vskip2pt\rightline{January, 1992.}
}
}
\vskip -1cm

\title{The Exact Renormalization Group and Approximate Solutions}
%\vbox{\centerline{
%}
%\vskip2pt\centerline{  }}
\vskip -1cm
\author{\bf Tim R. Morris}{}{}
%\myfoot{$^\dagger$}{\rm e-mail: trmorris@surya11.cern.ch}
\vskip 0.5cm
\addressline{\it CERN TH-Division}
\addressline{\it CH-1211 Geneva 23}
\addressline{\it Switzerland\myfoot{$^*$}{\rm On Leave from Southampton
University, U.K.}}
\addressline{\it }
\vskip -1cm

\abstract %130 words =14 lines
We investigate the structure of Polchinski's formulation of the flow
equations  for the continuum Wilson effective action.
  Reinterpretations in terms of I.R.
cutoff greens functions are given. A promising non-perturbative approximation
scheme is derived by carefully taking the sharp cutoff limit
 and expanding in `irrelevancy' of  operators.
 We illustrate with two simple models of four dimensional $\lambda \varphi^4$
theory: the cactus  approximation, and
 a model incorporating the first irrelevant correction to the
 renormalized coupling. The qualitative and quantitative behaviour give
confidence in a fuller use of this method for obtaining accurate results.

\vskip -1cm
%\draft
\date{\vbox{
{CERN-TH.6977/93}
\vskip2pt{SHEP 92/93-27}
\vskip2pt{hep-ph/9308265}
\vskip2pt{August, 1993.}
}
}

%%%%%%%%words and defs%%%%%%%%
\def\nonp{non-perturbative}
\def\pert{perturbative}
\def\DS{Dyson-Schwinger equations}
\def\etc{{\it etc.}\ }
\def\ie{{\it i.e.}\ }
\def\eg{{\it e.g.}\ }
\def\cf{{\it c.f.}\ }
\def\viz{{\it viz.}\ }
\def\pf{$\lambda \varphi^4$ theory}
\def\phi{\varphi}
\def\epsilon{\varepsilon}
\def\te#1{\theta_\epsilon( #1,\Lambda)}
\def\p{{\bf p}}
\def\P{{\bf P}}
\def\q{{\bf q}}
\def\r{{\bf r}}
\def\x{{\bf x}}
\def\y{{\bf y}}
\def\de#1{\delta_\epsilon( #1,\Lambda)}
\def\tr{{\rm tr}}
\def\Sv{S(\p_1,\cdots,\p_n;\Lambda)}
\def\D{{\cal D}}
\def\iprop{\Delta^{-1}\! }
\def\phil{\phi_<}
\def\phig{\phi_>}
\def\phic{\phi^c}
\def\phie{<\!\phi\!>}
\def\ipropl{\Delta_<^{-1}\! }
\def\ipropg{\Delta_>^{-1}\! }
\def\ins#1#2#3{\hskip #1cm \hbox{#3}\hskip #2cm}
%%%%%%%%%%%%%%%%%%%%%%%%%%%%%%

\section{Introduction.}
It need hardly be stated that an efficient   method of performing
accurate {\it ab initio}  \nonp\ {\sl continuum} calculations in
realistic quantum field theories would prove extremely useful.
 What is needed is to be able to produce a sequence of better and
better approximations in the sense that they can be seen to converge.
 The higher approximations must be  calculable
without inhuman effort, and if the method is to be  generally useful
it should apply even if there are no obviously
identifiable small parameters to control the approximation. This paper
we hope is a first step towards such a scheme.

The scheme indeed
seems to work, judging by the preliminary model approximations presented
in this paper. Since the majority of the paper discusses more formal aspects
we  limit ourselves here to these simple models, but the apologia should be
added immediately that these are not
enough to be sure that the requirements of convergence and calculability
really are satisfied. A proper demonstration of this requires that
 the method be applied to some real \nonp\ problem, such as triviality bounds
on the Higgs mass with and without\foot{to compare with lattice results}\ the
top.
This research is underway. One should also add that since the method
is tied to momentum cutoff, gauge-invariant theories present special
problems of their own.

We use Wilson's exact renormalization group\ref\kogwil{K. Wilson and J. Kogut,
Phys. Rep. 12C (1974) 75.}\  applied to the continuum field
theory with some physical cutoff, in the particularly simple form introduced by
Polchinski\ref\pol{J. Polchinski, Nucl. Phys. B231 (1984) 269.}. Our reasons
for using the exact renormalization group  are described in sect.2: it is the
only framework we know of which avoids the difficult U.V. problems afflicting
other approximation methods. In particular we contrast with  attempts to use
approximations to \DS , and briefly compare  other alternatives.

In the approach of ref.\pol\ the results are phrased in terms of an effective
action $S_\Lambda[\phi]$ with an intermediate-scale momentum cutoff $\Lambda$;
from this
we can construct the greens functions and S matrices of ultimate physical
interest but they will be restricted to having momenta less than $\Lambda$.
Originating with Keller, Kopper and Salmhofer\ref\kel{G. Keller, C. Kopper and
M. Salmhofer, Helv. Phys. Acta 65 (1992) 32} it was noticed that the same
equations as Polchinski's are also
satisfied essentially\foot{actually the external legs must be multiplied by
inverse bare propagators}\ by $W_\Lambda[J]$, the generator of connected greens
functions with I.R. momentum cutoff $\Lambda$. Therefore in this case as
$\Lambda\to 0$ one is left with greens functions defined for all physical
momenta.

It is immediately clear that the two constructs must be closely
related. While this is evidently understood by workers in the field,\foot{see
for example ref.\ref\salm{M. Salmhofer, Nucl. Phys. B (Proc. Suppl.) 30 (1993)
81}, ref.\ref\bonini{M. Bonini, M. D'Attanasio and G. Marchesini, Universit\`a
di Parma preprint UPRF 92-360}\ and references therein}\  no-one seems to have
precisely spelt out the relation. Therefore we start with this in sect.3. The
fact that,
as we make clear, the vertices of $S_\Lambda[\phi]$ are essentially both
vertices of the Wilson effective action
{\sl and}  I.R. cutoff connected greens functions, depending  on whether the
external momenta are all below or above $\Lambda$ respectively, proves very
useful both in our analytic investigations but also interpretationally.

We now  describe the results reported in the rest of the paper. We endeavour to
give here, as much as possible, an
intuitive explanation of the results; they are proved properly later. In some
places the reader might find the intuitive explanations unilluminating,
separated as they are from the details and exact definitions. If so the reader
should ignore them for the moment and return to them later, having read the
main body of the paper.
\nref\jf{J.F. Nicoll, T.S. Chang and H.E. Stanley, Phys. Rev. Lett. 33 (1974)
540. This is recalled in P. Hasenfratz and J. Nager, Z. Phys. C37 (1988)
477.}\nref\has{ A. Hasenfratz and P. Hasenfratz, Nucl. Phys. B270 (1986) 685}
\nref\wet{C. Wetterich, Phys. Lett. B301 (1993) 90,  and refs. therein.}
\nref\el{U. Ellwanger and L. Vergara, (1992) Heidelberg preprint
HD-THEP-92-49}\

Turning to the flow equations   we note that they are very sensitive to the
form
of the intermediate cutoff $\Lambda$,  even becoming
ambiguous in the limit of sharp cutoff. This is simply because they describe
the effect of integrating out momenta around the cutoff ($p\approx \Lambda$),
where the effective vertices are changing rapidly. Therefore, as the equations
stand,  they are inappropriate
for developing approximations. We resolve the limit  by finding the  general
solutions,
parametrized in terms of functions that are smoothly varying at $p=\Lambda$.
These functions can be chosen to coincide with the 1PI (one particle
irreducible) vertices of $\Gamma_\Lambda[\phi^c]$, the
Legendre effective action with I.R. momentum cutoff.
Intuitively this follows if one recalls that the vertices of $S_\Lambda$
correspond to I.R. cutoff connected greens functions. These have a tree
expansion
in 1PI vertices connected by I.R. cutoff full propagators. Thus $S_\Lambda$
can be reconstructed from the 1PI greens functions. But these latter are
smoothly varying at $p=\Lambda$
because in them all I.R. cutoff propagators are integrated over internal
momenta, so there are no terms left to vary rapidly as the external momenta
range over $p\approx\Lambda$.
Performing the Legendre transform
gives  flow equations for the 1PI vertices. (The flow equations are also
derived in refs.\wet\bonini. See also the ``Note Added'' at the end of the
paper.)  In fact we will show that the vertices of the
Legendre effective action $\Gamma_\Lambda$ are the same as those of the Wilson
effective action $S_\Lambda$ when all
partial sums of external momenta are less than $\Lambda$ (because in this range
  all the connecting I.R. cutoff  propagators in the tree expansion
vanish). Thus the dual interpretation
alluded to above is reflected here too. In fact $\Gamma_\Lambda[\phic]$ has
an interpretation as a {\sl Wilsonian quantum effective action} obtained by
integrating out purely quantum modes with momenta $p>\Lambda$; $S_\Lambda$
being obtained by integrating out, in addition, the tree level excitations
of the ``classical field'' $\phic$ with momenta $p>\Lambda$.

One final simplification of the structure of the equations is possible before
turning
to approximations. Up until now all manipulations have been regulated by an
overall
momentum cutoff $\Lambda_0$ and indeed this appears explicitly in the flow
equations.
One may ask however, why equations that only require momenta $p\approx\Lambda$
should
depend on $\Lambda_0>\! >\Lambda$ at all. In answering this question we show
that there
exists a reparametrization invariance of the flow equations that allows us
to change the cutoff $\Lambda_0$ without at all altering  $S_\Lambda[\phi]$.
The $\Lambda_0=\infty$ case gives the maximally analytic resolution
of the original flow equations. Solutions are obtained by setting boundary
conditions at some finite scale $\Lambda_0'$.
The reparametrization is simply a tree diagram expansion of the 1PI vertices
with overall momentum
cutoff $\Lambda_0$ in terms of the full propagators and 1PI vertices with a
higher overall momentum cutoff. Surprising as this result may seem it can be
understood intuitively from our previous comments: On the one hand the 1PI
vertices with the lower overall momentum cutoff are not 1PI with respect to
the higher overall momentum cutoff, because the expansion of connected greens
functions in terms of 1PI greens functions was performed only for propagators
with momenta less than the lower cutoff. On the other hand we know that we
may compute the Wilsonian effective action at cutoff $\Lambda=\Lambda_0$ in
terms of tree diagrams constructed from
the Legendre effective action with the higher overall momentum cutoff and I.R.
cutoff $\Lambda_0$. This Wilson effective action corresponds to the
bare action for an {\sl effective theory} with U.V. cutoff $\Lambda_0$. Using
this  to construct the {\sl effective} Legendre effective action by integrating
out quantum modes with momenta less than $\Lambda_0$ will give the latter a
tree expansion also.

Finally in sect.4 we discuss approximations. Needless to say the equations are
still not solvable exactly: there are an infinite set of them, they are
non-linear, and they
contain explicitly  I.R. cutoffs on all propagators. It is the last property
that makes
them  particularly difficult for analytic (and numerical) methods. For example
even in the sharp
cutoff limit, one-loop diagrams with external momentum dependence, are not
doable  exactly. It is therefore necessary that any approximation simplify the
effect of these cutoffs. We consider several approximations, but
the most promising is to make some expansion in the external momenta around
$\p={\bf0}$.  It turns out that a truncation coinciding with the simplest case,
where all external momentum dependence is discarded,\foot{See
ref.\wet\ref\wetter{
``Critical Exponents from the Effective Average Action'', N. Tetradis and C.
Wetterich, DESY and Heidelberg preprint (1993) DESY 93-094, HD-THEP-93-28}
 for an incorporation of one-loop
wavefunction renormalization.}\ has been discovered and rediscovered several
times\jf--\el\ (probably only a partial list, for further references see the
``Note Added'' at the end of the paper).
This truncates to an (uncontrolled) approximation
for the effective potential.
 Nevertheless the results are already encouraging, incorporating
as they do the exact one-loop answer plus some higher corrections, and give
reasonable
qualitative descriptions and numerical results for critical indices \etc in
scalar field theory for gaussian and non-gaussian fixed points and
dimensions $2\le D\le4$ \has\wetter .  The problem of sensitivity to the
form of the cutoff, discussed above and in sect. 3, can be avoided at this
level.

Considering now the higher orders, it becomes important whether one uses a
smooth\wet\el\ or sharp\jf\has\
cutoff. With a smooth cutoff $\te q$, momenta $\q$ in the flow equations are
integrated out
over a shell of radius $\Lambda$ and thickness $2\epsilon$. Evidently we must
have $\epsilon<\Lambda$. The momentum expansion corresponds to a local
derivative expansion in the effective lagrangian with a radius
of convergence $p\approx \epsilon$ (from expanding  terms such as
$\te{|\q+\p|}$ with $q\approx\Lambda$). Since these expansions are substituted
back in the flow
equations where they are themselves averaged over $p\approx\Lambda$, we must
have $\epsilon>\Lambda$ for convergence. The two inequalities together imply
that the expansion method converges
very weakly if at all.  One also
finds that the coefficients of the derivative terms are very sensitive
functions of the form of the cutoff\el , being proportional to integrals over
shells $q\approx\Lambda$  of higher derivatives of $\te q$ (which in general
must be performed numerically). This  makes their interpretation difficult.
Formally this sensitivity disappears in the limit $\Lambda\to0$ but also in
this limit the momentum expansion
breaks down completely.

For all these reasons  (explained further in sect.4) we choose instead to take
the sharp cutoff limit. Many integrals in the momentum expansion (averages over
3-spheres radius $q=\Lambda$) are now doable exactly, and the resulting flow
equation is a set of simultaneous first order differential equations which are
easily solved numerically. Moreover we expect the radius of convergence to be
the natural one (the minimum distance from the region $q>\Lambda$ to the first
singularity in complex $q$ space). However we no longer have the luxury of a
local effective lagrangian. The non-localities are non-analytic terms in the
momentum $p^\mu$ and are straightforwardly related to the fact that  both the
cutoff and momenta are now precisely determined: momentum modes $\phi(\q+\p)$,
where $p<\!<q=\Lambda$, are integrated out if and only if $\p.\q>0$, no matter
how small $p$ is. Thus the
expansion is in the length $p$ (not $\p^2$) and when several momenta are
involved the
coefficients are non-trivial functions of the angles subtended between them.

A sensible method of approximation is now to truncate at some order  $p^m$ and
at some $n$-point function (setting higher order in $n$ and $m$ arbitrarily to
zero), the levels
being determined by how relevant (in the colloquial sense) the higher
corrections are
to some physical quantity of interest. As a rule of thumb one might expect this
to
coincide with an expansion in increasing order of irrelevancy, as in the
technical sense.
We investigate in sect.5  two simple models, as a first check for  convergence,
calculability and any difficulties of interpretation in this scheme.

The first model is the analogue of the ladder (or cactus) approximation for
four dimensional \pf. This can be solved
analytically and the flows, fixed points \etc understood. We note that,
choosing a certain range of negative bare mass-squared $m_0^2$ and bare
coupling $\lambda_0$, and working in the unstable symmetric phase,
makes the model track towards the tachyon singularity, giving a peculiar
continuum limit. However we expect zero radius of convergence here. The problem
is cured by expanding about some other point $\phi=$ const.$\ne0$ \eg the true
minimum. Otherwise the results are all physically sensible. In particular we
find a sort of ``triviality of mass'', namely that it is impossible
to have a small renormalized mass-squared $m^2$ (compared to the cutoff
$\Lambda_0^2$) if the coupling $\lambda$ is greater than
a critical value $\lambda_c=8\pi^2\approx79$. This value, interpreted as the
maximum renormalized
coupling, agrees unreasonably well with previous precision lattice
calculations\ref\adel{M. L\"uscher and P. Weisz, Nucl. Phys. B290 (1987) 25.}:
$\lambda_{\rm max}=78\pm3$.

In the second model we investigate \nonp\ corrections to the coupling; the {\it
raison d'\^etre} of the method. Here we make the unnecessary ansatz, purely to
keep it technically simple, of setting all self-energies
to zero. Furthermore we truncate by setting the 6-point function to zero. The
perturbative
corrections to the coupling incorporate the exact one-loop result and  diagrams
at
two loops that are not associated with wavefunction or mass renormalization
(cf. fig.5).
The nested two-loop diagram is not calculable exactly (because of the I.R.
cutoffs) but
its contribution to the $\beta$-function is calculable to high accuracy as a
rapidly convergent {\sl numerical}
series, by expanding the inner loop in the outer loop's momenta. This is a nice
model confirmation
our general philosophy. Ordinarily such a method would fail
disastrously: the expansion being both increasingly I.R. and U.V.
divergent in the inner and outer loop respectively. It works here because inner
loop integrals are I.R. cutoff by the outer loop's
momentum. Performing the momentum expansion to next-to-leading order on the
\nonp\ equations, we construct a term of the form $\phi^2\sqrt{\lform}\ \phi^2$
in the effective lagrangian, giving a $\beta$-function which is an asymptotic
power series in $\lambda(\Lambda)$   incorporating the exact one-loop
contribution and  99.1\% of the 2-loop term. Integrating the \nonp\ equations,
they focus in on the gaussian fixed point: $\lambda(\Lambda)\to 0$ as
$\Lambda\to0$. The correction to the one-loop
result is small. For example in the ``scaling region''\adel ,
$\lambda(\Lambda)<\lambda_s\sim5$, we find that the percentage correction to
$\lambda(\Lambda)$ depends on $\lambda(\Lambda)$ as $a\lambda(\Lambda)$ where
$a=-1.7\%\ (-5.6\%)$ for bare couplings
$\lambda_0=20\ (1000)$. In fact there is a surprising amount of the physics of
triviality
captured by this model: one can show that the  \nonp\ effects are all cutoff
dependent and measure
their magnitude; compute the RG improved Symanzik terms, the large order
behaviour of the series and the renormalon ambiguity.
  Such investigations however
take us too far from the main purpose of this paper and will be reported on
elsewhere.

Taken together these preliminary tests suggest that we have the technique under
control,
while the smallness of the corrections to the one-loop case---even for very
large  bare couplings---gives confidence in a fuller use of these approximation
methods for obtaining accurate results.

\section{Renormalizability.}
As noted in the introduction our aim is to find a method of continuum
calculation in realistic quantum field theories consisting of a sequence of
better and better approximations, calculable without inhuman effort, and
applicable even if there are no obviously
identifiable small parameters to control the approximation.

Our first thought in this direction was to use the \DS . It is well known
that these are solvable in the large $N$ approximation. However one quickly
runs up against the  above desiderata: while higher corrections in realistic
cases, \eg the
Higgs sector  with $O(4)\to O(N)$, are calculable in principle, in practice the
difficulties are such that the results are essentially restricted
to $N=\infty$\ref\largen{ For example, ``Large N Analysis Of The Higgs Mass
Triviality Bound'',
 Urs M. Heller, Herbert Neuberger, Pavlos Vrana, Florida State University
preprint FSU-SCRI-92-99, Jul
1992.},
while it is easy to envisage cases where the true
$N$ is definitely not large enough  for the asymptotic expansion in $1/N$
to give good estimates to any order. Instead we tried to  justify and improve
the ``ladder\foot{{\it a.k.a.} ``rainbow'' or (for \pf) ``cactus''}\
approximation'' and its (arguably better) variants\ref\pen{ ``Nonperturbative
Study of the Fermion Propagator in Quenched QED in covariant gauges using a
renormalizable truncation of the Schwinger-Dyson Equation'', D.C. Curtis and
M.R. Pennington, Durham preprint (1993) DTP-93-20 and refs. therein.},\foot{got
by keeping only the  \DS\ for the self-energy and introducing some ansatz for
the vertex}\  by incorporating directly some more
of the infinite set of \DS . It is here that one runs into what appears to be
an insuperable problem: In order to make progress one
must truncate the higher equations somehow, but the results (beyond the ladder)
seem always not to be  even
perturbatively renormalizable. This is because one throws away with the
truncation certain sets of  divergent diagrams, but {\sl keeps}
the simpler counterterm diagrams generated by lower terms in the \DS . This
problem is only seen beyond the ladder because it is only here that the
coupling constant starts receiving divergent corrections to all orders in
perturbation theory. The counterterms to which we refer are
those generated by replacing the bare vertices in the lower terms of the \DS\
by the required power series in renormalized coupling. While these cancel
divergences in diagrams that are kept, they also
generate counterterm diagrams corresponding to diagrams that have been thrown
away with the truncation.
Since the counterterms no longer have divergences to cancel, they contribute
divergent terms (in particular some of the
form $\ln p\ln\Lambda_0 $ \etc$\!$) which destroy perturbative
renormalizability. It is not clear to us that this problem can at all
be solved, but more importantly even if it can,  the resulting sequence of
approximations are unlikely to be easily calculable.

(A number of authors have tried to improve the ladder ansatz by  perturbatively
correcting the kernels, \eg  the photon propagator in strong coupling QED.
This only worsens the approximation however because the perturbative series is
asymptotic and inapplicable to the strong coupling regime. It is easy to
explicitly verify this with a simple model ($0+0$ dimensional \pf) but this
would take us too far from the main thrust of this paper.)

This experience teaches us an important lesson however, which is applicable
to many a  scheme involving truncations\ref\tamm{See for example
``Renormalization of Tamm-Dancoff Integral Equations'' Brett van de Sande and
Stephen S. Pinsky, Phys.Rev.D46 (1992) 5479}: namely that perturbative
renormalizability is {\sl not} guaranteed. If it is truly necessary to test all
attempts to repair or improve
 these  schemes by some analogue of the detailed graphical and convergence
arguments
 used in the classic proofs of perturbative renormalizability, then this is
rather depressing, not the least because they likely fail the test and in
any case {\sl\nonp }\ renormalizability---even for very small coupling---would
remain
an open question.

It is for the above reasons that we turned to the exact renormalization group
as the appropriate framework for \nonp\ approximations: As pointed out by
Polchinski in a beautiful paper\pol , in the Wilson renormalization
group framework the perturbative renormalizability of \eg four dimensional \pf,
is essentially obvious. It is only necessary to show that
the operators we know are irrelevant (\ie have negative scaling dimension) at
zero coupling, remain so at very small
coupling. But this is so because the right hand side of the Wilson flow
equations $\partial S_\Lambda/\partial\Lambda=\cdots$, where $\Lambda$ is an
intermediate cutoff for the effective action $S_\Lambda$, is a smooth function
of the coupling. Indeed the detailed proof\pol\ only requires very simple
bounds on the right hand side which do not involve any cancellations between
terms, and thus it is clear, both intuitively and
in detail, that {\sl truncations} of the flow equations are also perturbatively
renormalizable. Of course what happens for large couplings is a different
matter since here renormalizability even in the full theory is brought into
question, but at least one can be sure that this framework is
sensible in the small coupling regime.

Actually the problem of renormalizability in truncations has its analogue in
difficulties
with any approximation, in the following sense: If we are only modelling the
high energy behaviour approximately then the divergences of quantum field
theory will tend to ensure
that the approximation is infinitely bad in the limit that the cutoff is
removed. It is
such an unwelcome sensitivity to high energies (together with the inability to
compute any but the simplest approximation) which is the main reason for the
failure of variational methods in quantum field theory\ref\fey{R. P. Feynman,
``Difficulties in Applying the Variational Principle to Quantum Field
Theories'' in ``Variational Calculations in Quantum Field Theory'' Wangerooge
proceedings (1987), World Sci. Eds. L. Polley and D.E.L. Pottinger}. Lattice
methods suffer too in the sense that  nearly  all the numerical effort is
expended
on computing effects of order the lattice spacing, the time required and the
number  of lattice points required  diverging as the parameters are tuned
to the continuum limit.

The methods described in this paper appear not to suffer from
this general problem;  on the one hand the calculations are protected
from U.V. divergences by the intermediate cutoff $\Lambda$ and on the other,
even approximations to the exact
renormalization group  focus the effective lagrangian ---
erasing errors  in irrelevant degrees of freedom
and automatically absorbing them in the low energy definition of relevant and
marginal couplings.

\section{The Flow Equations and Their Structure.}
Throughout the paper we work with a one-component scalar field $\phi(\x)$: the
generalisation to more components and fermions is straightforward. As usual in
momentum space, we factor out and evaluate the momentum conserving
$\delta$-functions so $n$-point greens functions are written as $\sim$
$G(\p_1,\cdots,\p_n)$ and
are defined only when $\p_1+\cdots+\p_n={\bf0}$, while in addition in two-point
functions we solve $\p_1=-\p_2=\p$ and recognise that they are functions only
of $p=|\p|$ \eg the propagator is written $\Delta(p)$. We use the condensed
notation wherever convenient: $\phi.J\equiv\phi_x J_x\equiv \int\! d^Dx \
\phi(\x)J(\x)$ where
$D$ is the dimension of euclidean space-time. Similarly the propagator
$\Delta(\x,\y)$, and other greens functions of two arguments, will be regarded
as a matrix so $\phi.\Delta^{-1}\!.\phi$
means $\phi_x(\Delta^{-1})_{xy}\phi_y\equiv\int\! d^Dp/(2\pi)^D \
\phi(\p)\left[\Delta(p)\right]^{-1}\phi(-\p)$. The partition function is
assumed to be regulated by an overall momentum cutoff $\Lambda_0$; this
will be made explicit in a moment.

Let us start by observing that the partition function for the scalar field
$\phi$ with propagator $\Delta$ and arbitrary bare interaction
$S_{\Lambda_0}[\phi]$,
\eqn\zorig{Z[J]=\int\!\D\phi\
\exp\{-\half\phi.\iprop.\phi-S_{\Lambda_0}[\phi]+J.\phi\}}
can be rewritten in terms of two propagators and two fields as
\eqn\fact{\eqalign{Z[J]=\int\!\D\phig\D\phil\
 \exp\{-\half\phig.\ipropg.\phig &-\half\phil.\ipropl.\phil \cr
 &-S_{\Lambda_0}[\phig+\phil]+J.(\phig+\phil)\}\cr }}
(up to a multiplicative factor which we always ignore), where
\eqn\summ{\Delta=\Delta_<+\Delta_>\hskip 1cm {\rm and} \hskip 1cm
\phi=\phig+\phil\ \ .}
This is true whatever `partition' of $\Delta$ is used and is obvious in
perturbation theory,  since every propagator in each graph is just repeated
twice ---
once with $\Delta_<$ and once with $\Delta_>$, but it is
also true \nonp ly. (Substitute $\phig=\phi-\phil$ followed by
$\phil=\phil'+(\Delta_</\Delta).\phi$, and integrate out the now gaussian
$\phil'$).
This trick was used in ref.\ref\zin{ J. Zinn-Justin, ``Quantum Field Theory and
Critical Phenomena'' (1989) Clarendon Press, Oxford.}\ for a nice constructive
proof of Polchinski's flow equations. Here we will use it to further
investigate
the properties of the effective action.

 Write
\eqn\deltas{\Delta_>(p)=\left[\te p -\theta_\epsilon(p,\Lambda_0)\right]\
\Delta(p) \ins11{and}\Delta_<(p)=\left[1-\te p\right]\ \Delta(p)\quad\quad,}
where $\te p$ is a smooth cutoff function, bounded above (below) by
one (zero), satisfying
$\te p\approx 0$ for $p<\Lambda-\epsilon$ and $\te p\approx1$ for
$p>\Lambda+\epsilon$. Thus $\Delta_>(p)$ is the propagator cut off
from below by $\Lambda$ and above by $\Lambda_0$, while $\Delta_<(p)$ is the
propagator cut off from
above by $\Lambda$. We have taken the opportunity here to make explicit the
overall momentum cutoff by replacing $\Delta(p)$ in \zorig--\summ\ by $\left[
1-\theta_\epsilon(p,\Lambda_0)\right] \Delta(p)$. The intention eventually is
take
the sharp cutoff limit:
\eqn\thelimit{\te p\to \theta(p-\Lambda)\ins11{as} \epsilon\to0}
and thus to identify, by \fact, $\phil(\p)$ with the low momentum modes
$p<\Lambda$ and $\phig(\p)$ with high momentum modes $p>\Lambda$.
But  we keep
$\Delta_<(p)$ and $\Delta_>(p)$ non-zero for all $\p$ at intermediate stages,
to
 inject an element of rigour into the approach. (Observe that for limit
\thelimit\ the inverses in \fact\ are not defined).

Integrating out the nascent high momentum modes $\phig$ in \fact\ we have
\eqn\zless{Z[J]=\int\!\D\phil\ \exp\{-\half\phil.\ipropl.\phil
-S_\Lambda[\Delta_>.J+\phi_<]+J.\phil
+\half J.\Delta_>.J\}}
for some functional $S_\Lambda$. To see this, isolate in \fact\ the integral
over $\phig$ dependent factors, and substitute $\phig=\phi-\phil$:
\eqnn\ztozl
\eqnn\zl
\eqnn\zlphi
$$\eqalignno{&Z[J] = \int\!\D\phil\ \exp\{-\half\phil.\ipropl.\phil\}
\  Z_\Lambda[\phil,J]\ins{1.5}{1.5}{where}&\ztozl\cr
&Z_\Lambda[\phil,J]=\int\!\D\phig\
\exp\{-\half\phig.\ipropg.\phig-S_{\Lambda_0}[\phig+\phil]
+J.(\phig+\phil)\} &\zl\cr
=&\exp\{-\half\phil.\ipropg.\phil\}\int\!\D\phi\
\exp\{-\half\phi.\ipropg.\phi-S_{\Lambda_0}[\phi]+\phi.(J+\ipropg.\phil)\}\ .
&\zlphi\cr}$$
 Integrating over $\phi$ gives
$$\eqalign{Z_\Lambda[\phil,J]
=\exp\{\half J.\Delta_>. J+J.\phi_<\}\
\Bigr\{\exp-&\half(J  +\ipropg.\phil).\Delta_>.(J+\ipropg.\phil)\ \times\cr
\exp-S_{\Lambda_0}\![{\delta\over\delta J}]\
 \exp&\half(J+\ipropg.\phil).\Delta_>.(J+\ipropg.\phil)\Bigr\}
\ .\cr}$$
 Performing all the derivatives in $S_{\Lambda_0}[{\delta\over\delta J}]$ and
noting that
 the ${\delta\over\delta J}$'s are replaced by either $\Delta_>.J+\phi_<$ or
the differential of this ($\Delta_>$), we have for some $S_\Lambda$
\eqn\zlresult{
Z_\Lambda[\phil,J]=\exp\{\half J.\Delta_>.J+J.\phi_<
-S_\Lambda[\Delta_>.J+\phi_<] \}\quad,}
proving the assertion.

Now consider the limit \thelimit . If we also insist that  $J(\p)=0$ for
$p>\Lambda$, so that $J$ only couples to the low momentum modes
 (as in ref.\pol), we see
that  in \zless\ all the $J$'s drop out except for $J.\phil$, and thus
$S_\Lambda$
coincides with the Wilsonian effective action\foot{in the same form as ref.\pol
}: it is the same form as \zorig\ with $\Lambda_0\mapsto\Lambda$. We can keep
this interpretation for general $J(\p)$ if we interpret $J$ in \zorig\ as a
space-time dependent
one-point coupling, analogous to a non-constant external magnetic field.

On the other hand
if we set $\phil\equiv0$ in \zl , we have a standard partition function for a
field $\phig$
 with an I.R. cutoff $\Lambda$ imposed. Thus using \zlresult\ and defining
\eqn\defW{W_\Lambda[\phil,J]=\ln Z_\Lambda[\phil,J]\ =\half
 J.\Delta_>.J+J.\phi_< -S_\Lambda[\Delta_>.J+\phi_<] \quad,}
we have that $W_\Lambda[0,J]=\half J.\Delta_>.J-S_\Lambda[\Delta_>.J]$ is the
generator of connected greens functions with I.R. cutoff $\Lambda$. This is the
relation advertised in the introduction: we see that in the limit \thelimit\
the support of $S_\Lambda$ neatly separates into low
momenta (provided by $\phil$) where it is the Wilsonian effective action,
and high momenta (provided by $\Delta_>.J$) where it is
related to the generator of connected greens functions as above. To make
the relation completely explicit, define $\Phi=\Delta_>.J+\phil$ and
\eqn\defS{{S(\p_1,\cdots,\p_n;\Lambda)={\delta^n S_\Lambda[\Phi]
\over \delta\Phi(\p_1)\cdots\delta\Phi(\p_n)}\quad .}}
Writing the I.R. cutoff connected greens functions as
$$G(\p_1,\cdots,\p_n;\Lambda)={\delta^n W_\Lambda[0,J]\over\delta J(\p_1)
\cdots\delta J(\p_n)}\quad, $$
we obtain in the limit \thelimit\
\eqn\vertexGS{\eqalign{
\Lambda > {\rm all}\ p_i\quad : \quad S(\p_1,\cdots,\p_n;\Lambda) &=
\hbox{vertex of effective action at momentum}\cr
&\phantom{\hbox{vertex of effective action at moment}}{\rm scale}\ \Lambda\cr
\Lambda < {\rm all}\ p_i\quad :\quad S(\p_1,\cdots,\p_n;\Lambda) &=
-G(\p_1,\cdots,\p_n;\Lambda)\prod_{i=1}^n \Delta^{-1}(p_i)
%\cr
%&\hskip 6cm
\ {\rm for}\ n>2\cr
S(p;\Lambda) &={1\over\Delta^2(p)}[\Delta(p)-G(p;\Lambda)]
\hskip 1.5cm {\rm at}\ n=2\cr
}}
(Of course in the second relation we need also all $p_i<\Lambda_0$).

At  this stage we interrupt the exposition with some pedantry. Firstly observe
that, in the limit \thelimit, $\phil(\p)=0$ for $p>\Lambda$
is not imposed in \zlresult, but only fluctuations satisfying this
give non-zero contributions to \ztozl.  Our later proofs can be shortened by
using this
observation to note that $S_\Lambda$ in \zlresult\ may therefore be
regarded as
over-parametrized. Indeed all that is needed
to identify the Wilsonian effective action as above, and effectively to make
 all our later observations also, is the limit
$\Delta_>(p)\to0$ (as $\epsilon\to0$) for $p<\Lambda$. Thus
(in the limit) we require $\Delta_<(p)=\Delta(p)$  for the low momentum modes
$p<\Lambda$, while for $p>\Lambda$ the propagator $\Delta_<(p)$
can for example have a high energy `tail', giving $\phil$ support
on all momenta. We will not continue to make or use such comments,
because they neither clarify nor are they important
for our present purposes. The same  applies to
modifications of our observations to fit the  appropriate inequalities
and approximate relations for a smooth cutoff \deltas, and
 where the approximate relations become exact for $\te p$ an exact partition of
unity. We trust the reader interested in these cases
will find it straightforward to modify the proofs as appropriate.

Let us define
\eqn\defK{K_\Lambda(p)=-{d\over d\Lambda}\Delta_>(p)=\de p\ \Delta(p)}
so, by \thelimit ,
\eqn\defde{\de p \equiv -{d\over d\Lambda}\te p\to\delta(p-\Lambda) \ins11{as}
\epsilon\to0\quad.}
{}From the definition \zl\ we have
\eqn\flowZ{{d\over d\Lambda} Z_\Lambda[\phil,J] =-{1\over2}
\left({\delta\over\delta J}-\phil\right).\left({d\over
d\Lambda}\ipropg\right).\left({\delta\over\delta J}-\phil\right)\
Z_\Lambda\quad,}
which, on substituting \zlresult,  gives
\eqn\Peqn{{\partial\over\partial\Lambda}S_\Lambda[\Phi]=
{1\over2}\left\{ {\delta S_\Lambda\over\delta\Phi}.K_\Lambda.{\delta
S_\Lambda\over\delta\Phi} - \tr\left(K_\Lambda.{\delta^2
S_\Lambda\over\delta\Phi\delta\Phi}\right)\right\}\quad .}
({\it N.B.} again, in this equation $K_\Lambda$ and the  $\Phi$ two-point
function are regarded
as  matrices). This the Polchinski equation\pol .  It can also be
written in linear form as
$${\partial\over\partial\Lambda}\exp-S_\Lambda[\Phi]=
-{1\over2}\tr\left(K_\Lambda.{\delta^2
\over\delta\Phi\delta\Phi}\exp-S_\Lambda[\Phi]\right)\quad,$$
but this doesn't seem to be useful. Expanding \Peqn\ using \defS,
we obtain
\eqn\Pexpanded{\eqalign{
{\partial\over\partial\Lambda}S(\p_1,\cdots,\p_n;\Lambda)=
\sum_{\{I_1,I_2\}} &S(-\P_1,I_1;\Lambda)K_\Lambda(P_1)S(\P_1,I_2;\Lambda)\cr
&-{1\over2}\int\!{d^Dq\over(2\pi)^D}\,
K_\Lambda(q)S(\q,-\q,\p_1,\cdots,\p_n;\Lambda)\ .}
}
In here $I_1$ and $I_2$ are disjoint subsets ($I_1\cap I_2=\emptyset$) of the
momenta such that $I_1\cup I_2=\{\p_1,\cdots,\p_n\}$. The sum over
$\{I_1,I_2\}$ means a
sum over all such disjoint subsets, but pairs are counted only
once  \ie $\{I_1,I_2\}\equiv
\{I_2,I_1\}$, utilising the Bose symmetry of the equation. The momentum $\P_1$
is given by momentum conservation
as $\P_1=\sum_{\p_i\in I_1}\p_i$\ . The equation is best appreciated
graphically ---\cf fig.\fig\Pexpfig{ }.
\midinsert
\centerline{
\psfig{figure=exactrgfig1.ps,width=7in}}
\bigskip

\centerline{\vbox{{\bf Fig.1.} The Polchinski equation for the vertices.
These are drawn as open circles, while the two-point function $K_\Lambda$ is
drawn as a black dot.}}
\endinsert
Eqn.\Pexpanded\ gives the change in the vertices of the effective action, under
an infinitessimal lowering of the  intermediate cutoff $\Lambda$, as an
infinite
set of simultaneous first-order differential equations in $\Lambda$. Evidently
they can  in principle be used to define and solve the theory,
the initial boundary conditions being given by $S_\Lambda=S_{\Lambda_0}$
at $\Lambda=\Lambda_0$. With just two terms, totally symmetrized, the first
representing
tree-level contractions between vertices with the same or smaller numbers
of legs, and the second a one-loop integral over a vertex with
two more legs, it is about as simple as one could hope for.

They
are not however  simple from the point of view of approximations. The problem
is
that the R.H.S. (right hand side) of \Pexpanded\ has momenta $P_1$ and $q$
effectively restricted to the range $\Lambda-\epsilon<P_1,q<\Lambda+\epsilon$
(\cf  \defK\ and discussion
below \deltas), but either from \Pexpanded\ directly or by thinking
about the correspondence \vertexGS\ we see that the vertices of
$S_\Lambda$ sensitively depend on the form of $\theta_\epsilon$ there.  (It is
also helpful to think about this perturbatively \ie in terms
of connected Feynman diagrams using \defW: the $\theta_\epsilon$'s
appear in every propagator). Of course the exact $S(\p_1,\cdots,\p_n;\Lambda)$
is not sensitive to the form of $\theta_\epsilon$ for momenta $p_i$ well away
from $\Lambda$ (and $\Lambda_0$), but it should be clear that if we attempt to
approximate equations \Pexpanded\ directly, by truncation and/or otherwise,
this property of the equations will ensure in general that
our `approximate' solutions depend very sensitively on the choice
of $\theta_\epsilon$ and truncation \etc The problem becomes
especially clear if we consider the limit \thelimit. Then the question is how
to take the limit of the R.H.S.
of \Pexpanded\ remembering that $S(-\P_1,I_1;\Lambda)$ for example
is an as-yet unknown function of $\te{P_1}$. The answer is not at all
the usual physicist's expedient of putting $\theta(0)=\half$.

To demonstrate this, let us prove a little lemma:
\eqn\lemma{\de p f(\te
p,\Lambda)\to\delta(\Lambda-p)\int_0^1\!\!dt\,f(t,p)\ins11{as}\epsilon\to0}
where $f(\theta_\epsilon,\Lambda)$ is any function whose dependence
on the second argument ($\Lambda$) remains continuous at $\Lambda=p$
in the limit $\epsilon=0$. This follows from the identity
$$\de p f(\te p,\Lambda)=\left\{{\partial\over\partial\Lambda}
\int^1_{\te p}\!dt\, f(t,\Lambda')\right\}\Biggr|_{\Lambda'=\Lambda}$$
(\cf \defde), by noting that (by the properties below \deltas )
the integral is  a representation of a step-function in $\Lambda$ but with
height $\int_0^1\!dt\,f(t,\Lambda')$. As a consequence we have
for example $\de p\te p\to\half\delta(\Lambda-p)$ as expected,
but $\de p\theta^2_\epsilon(p,\Lambda)\to{1\over3}\delta(\Lambda-p)$.

It follows then that we need to know precisely the dependence of
the $S$'s on $\te {p_i}$, where $\p_i$ is any external momentum,  in order to
take the limit. On the other hand if we do not take the
limit then we must be careful in approximations to accurately
describe the dependence of the $S$'s on the $\te {p_i}$. For these reasons we
need to resolve the limit
by finding  general solutions for $\Sv$ parametrized in terms
of functions that are continuous at $p_i=\Lambda$.

Actually, finding such general solutions is straightforward.
For simplicities sake we specialize to the case that $S_{\Lambda_0}$ has only
even powers of $\phi$, so that the ${\rm Z}_2$ symmetry
$\phi\leftrightarrow-\phi$ ensures  the $\Sv$ with odd $n$ vanish. We
also take $\iprop\,(p)=p^2$ in \zorig,
putting the bare mass term $\half m_0^2\phi^2$ in $S_{\Lambda_0}[\phi]$. (Of
course the results
are independent of this split as we will shortly see: we choose it only because
it makes
the equations a little neater).

 Consider
first the equation for the two-point function. From \Pexpanded\ this
is
\eqnn\twoP
$$\eqalignno{
{\partial\over\partial\Lambda}S(p;\Lambda)&=
 K_\Lambda(p)S^2(p;\Lambda)
-{1\over2}\int\!{d^Dq\over(2\pi)^D}\, K_\Lambda(q)S(\q,-\q,\p,-\p;\Lambda)\cr
\hbox{with b.c.}\qquad S(p;\Lambda_0)&=m_0^2\quad,&\twoP\cr}$$
 where b.c. means boundary condition.  This has solutions $S$ that,
while discontinuous at $\Lambda=p$ in the limit $\epsilon=0$, neither vanish
nor diverge there. For such solutions
$$-{\partial\over\partial\Lambda}\left({1\over S(p;\Lambda)}\right)
=K_\Lambda(p)+\ finite\quad,$$
where $finite$ means a term that is finite at $\Lambda=p$ in the limit
$\epsilon=0$. Integrating,
using \defK\defde, we thus have
\eqn\defSigma{{1\over S(p;\Lambda)} ={1\over p^2}[\te p
-\theta_\epsilon(p,\Lambda_0)] +{1\over\Sigma(p;\Lambda)}}
where $\Sigma(p;\Lambda)$ is continuous and non-vanishing at $\Lambda=p$, in
the
limit $\epsilon=0$,  and satisfies the b.c. $\Sigma(p;\Lambda_0)=m_0^2$.
$\Sigma(p;\Lambda)$ has the interpretation of an  effective
self energy, indeed in the limit $\epsilon=0$ we
have
\eqn\interpStwo{
\eqalign{S(p;\Lambda)&=\Sigma(p;\Lambda)\ins21{for} p<\Lambda,\cr \hbox{while
from \vertexGS\ we have}\qquad G(p;\Lambda)&={1\over
p^2+\Sigma(p;\Lambda)}\ins1{0}{for}\ \Lambda_0>p>\Lambda.\cr}}
The latter equation  also implies that $\Sigma$ is 1PI (one particle
irreducible).

Now consider the equation for the 4-point function. From \Pexpanded\ this is
$$\eqalignno{{\partial\over\partial\Lambda}S(\p_1,\p_2,\p_3,\p_4;\Lambda)&=
S(\p_1,\p_2,\p_3,\p_4;\Lambda)\sum_{i=1}^4 K_\Lambda(p_i)S(p_i;\Lambda) \cr
&-{1\over2}\int\!{d^Dq\over(2\pi)^D}\,
K_\Lambda(q)S(\q,-\q,\p_1,\p_2,\p_3,\p_4;\Lambda)\cr
\noalign{\rm thus}\cr
{\partial\over\partial\Lambda}\ln S(\p_1,\p_2,\p_3,\p_4;\Lambda)&=
\sum_{i=1}^4 K_\Lambda(p_i)S(p_i;\Lambda) +finite\cr
&=\sum_{i=1}^4{\partial\over\partial\Lambda}\ln S(p_i;\Lambda)
+ finite\quad.\cr
}$$
where in the last line we used \twoP , and $finite$ now means a term
that is finite at $\Lambda=p_i$ \ $i=1,\cdots,4$ in the limit $\epsilon=0$.
Integrating up we have
\eqn\StoGam{S(\p_1,\p_2,\p_3,\p_4;\Lambda) =\Gamma(\p_1,\p_2,\p_3,\p_4;\Lambda)
\prod_{i=1}^4{S(p_i;\Lambda)\over\Sigma(p_i;\Lambda)}}
for some 4-point function $\Gamma(\p_1,\p_2,\p_3,\p_4;\Lambda)$ which is
continuous, in the limit $\epsilon=0$, at the
points $\Lambda=p_i$ \ $i=1,2,3,4$. The product over
$\Sigma$'s is allowed because it also has this property. It was slipped in
because it gives $\Gamma$ a neat interpretation, namely
it is the 4-point vertex of both the Wilsonian effective action
and the Legendre effective action (and thus 1PI). Indeed  in the limit
$\epsilon=0$ we have,  using \vertexGS,\defSigma\ and \interpStwo:
$$\eqalign{
S(\p_1,\p_2,\p_3,\p_4;\Lambda)&=\Gamma(\p_1,\p_2,\p_3,\p_4;\Lambda)
\phantom{-\prod_{i=1}^4 G(p_i;\Lambda)}\ins1{0.5}{all} p_i<\Lambda\cr
G(\p_1,\p_2,\p_3,\p_4;\Lambda)&=-\Gamma(\p_1,\p_2,\p_3,\p_4;\Lambda)
\prod_{i=1}^4 G(p_i;\Lambda)\ins1{0.5}{all} p_i>\Lambda\quad .\cr}$$

At this stage we take the hint that, by judicious parameterisation,
the continuous (in fact smooth) parts of $\Sv$ can be identified
with the 1PI (one particle irreducible) vertices of the Legendre
effective action. From the structure of the flow eqns \Pexpanded, we see that
$\Sv$ has an expansion in tree diagrams
 which on any 1PR (one particle reducible) leg, carrying momentum
$\sum_{\p_i\in I}\p_i$ where $I$ is a subset of the momenta
$\{\p_1,\cdots,\p_n\}$, there is a term containing $\te{|\sum_{\p_i\in
I}\p_i|}$: it is this that gives the first term in \Pexpanded . Indeed
it is obvious diagrammatically, through the identification below \defW, that
$\Sv$ is made up of 1PI bits which are continuous in the limit (since
all $\theta_\epsilon$'s are integrated over), connected by 1PR
legs through terms containing the factor $\Delta_>(|\sum_{\p_i\in I}\p_i|)$. We
thus have immediately that, in the limit $\epsilon=0$, only purely 1PI
contributions remain if  $\Lambda>$ all $|\sum_{\p_i\in I}\p_i|$ \ie from
\vertexGS:
\eqn\vertexSGam{S(\p_1,\cdots,\p_n;\Lambda)
=\Gamma(\p_1,\cdots,\p_n;\Lambda)\ins{0.3}{0.3}{if} \Lambda>|\sum_{\p_i\in
I}\p_i|\ins{0.3}{0.2}{for all} I\subset \{\p_1,\cdots,\p_n\},}
where $\Gamma(\p_1,\cdots,\p_n;\Lambda)$ are the 1PI vertices of
the Legendre effective action for the theory with I.R. cutoff $\Lambda$.
(Compare \interpStwo\ and the previous eqn.).

Now let us give a full \nonp\ proof of these assertions.
First we note that by \zlphi,
 $W_\Lambda[\phil,J]=\ln Z_\Lambda[\phil,J]$
 is in fact the generator of connected greens functions
even for $\phil\ne0$. Thus $\Gamma_\Lambda$ defined by
\eqn\defGam{\half(\phic-\phil).\ipropg.(\phic-\phil)
+\Gamma_\Lambda[\phic]=-W_\Lambda[\phil,J]+J.\phic}
is, by the usual analysis, the generator of 1PI greens functions.
In here $\phic$ is the classical field $\phic=\delta W_\Lambda\big/
\delta J$.
%\eqnn\defphic
%\eqnn\eqmo
%$$\eqalignno{\phic&={\delta W_\Lambda\over\delta J}\quad. %&\defphic\cr
%\hbox{We also have}\qquad {\delta \Gamma_\Lambda\over\delta\phic}
%+\ipropg.(\phic-\phil) &=J\quad. &\eqmo \cr}$$
 The extra terms on the left hand side are present in
the classical action (\cf \zlphi) and would normally be included
in the Legendre effective action but are not strictly speaking 1PI diagrams,
therefore we have written them separately. The reader may wonder
if we are being sloppy here by not displaying the dependence of
$\Gamma_\Lambda$ on $\phil$. In fact it is clear
that $\Gamma_\Lambda$ has no perturbative dependence on $\phil$, since using
\zlphi\ one cannot draw 1PI Feynman diagrams connecting $\phil$.
Non-perturbatively it is true too since we have from the definition \defGam:
$${\delta\Gamma_\Lambda[\phic]\over\delta\phil}\Biggr|_{\phic}
=\ipropg.(\phic-\phil)-{\delta W_\Lambda\over\delta\phil}\Biggr|_J$$
but the R.H.S. vanishes, since from \defW\ one proves
$${\delta W_\Lambda\over\delta\phil}=\ipropg.\left(
{\delta W_\Lambda\over\delta J}-\phil\right)\quad .$$
Substituting \defW\ into \defGam, and rearranging to get
$\Phi=\Delta_>.J+\phil$, one finds the following generalised Legendre transform
relation between $S_\Lambda$ and $\Gamma_\Lambda$:
\eqn\SG{
S_\Lambda[\Phi]=\Gamma_\Lambda[\phic]+\half(\phic-\Phi).\ipropg.(\phic-\Phi)}
which in particular implies $\phic=\Phi-\Delta_>.(\delta
S_\Lambda\big/\delta\Phi)$. Substituting this back into \SG\ gives:
\eqn\GamtoS{S_\Lambda[\Phi]=\Gamma_\Lambda[\Phi-\Delta_>.{\delta
S_\Lambda\over\delta\Phi}]\ +{1\over2}{\delta
S_\Lambda\over\delta\Phi}.\Delta_>.{\delta S_\Lambda\over\delta\Phi}\quad .}
By iteration, starting with $S_\Lambda[\Phi]=\Gamma_\Lambda[\Phi]$, one obtains
the now expected
expansion of $S_\Lambda$ in tree diagrams (as follows from \defW\ and described
above \vertexSGam). On the other hand for the vertex $\Sv$, if we ensure
that $\Lambda>|\sum_{\p_i\in I}\p_i|$ for all subsets $I\subset
\{\p_1,\cdots,\p_n\}$  and take the limit $\epsilon\to0$, then the above
$\Delta_>$ terms all vanish, proving \vertexSGam . Physically,
recalling that tree diagrams correspond to a small field expansion
of classical field theory, we may interpret result \GamtoS\
as follows: $\Gamma_\Lambda$ is the {\sl Wilsonian quantum effective
action} obtained from integrating out purely quantum field excitations with
invariant momentum $p>\Lambda$, while in $S_\Lambda$
in addition we integrate out also classical field excitations with
momentum $p>\Lambda$. For sufficiently small external momenta a given tree
level process with intermediate momentum $p>\Lambda$ is not possible, so here
the vertices of the two effective actions coincide.

We now derive the flow equations for $\Gamma_\Lambda$, from which we will see
explicitly that the dependence on $\Lambda$ is smooth, and the limit
$\epsilon=0$ may be taken unambiguously. Substituting
\defW\ into \flowZ\ we have
$${\partial W_\Lambda\over\partial\Lambda}\Biggr|_J=
-{1\over2}(\phic-\phil).{d\ipropg\,\over d\Lambda}.(\phic-\phil)
-{1\over2}\tr\left({d\ipropg\,\over d\Lambda}.{\delta^2 W_\Lambda\over
\delta J\delta J}\right)\quad,$$
which, using \defGam\ and the relation
$${\delta^2 W_\Lambda\over
\delta J\delta J}=\left(\ipropg\, + {\delta^2 \Gamma_\Lambda\over
\delta\phic\delta\phic}\right)^{-1}$$
(derived in the standard way from \defGam), gives
\eqn\thestart{
{\partial\over\partial\Lambda}\Gamma_\Lambda[\phic]\Biggr|_{\phic}
={1\over2}\tr\left\{{d\ipropg\,\over d\Lambda}.
\left(\ipropg\, + {\delta^2 \Gamma_\Lambda\over
\delta\phic\delta\phic}\right)^{-1}\right\}\quad.}
We now drop the uninteresting field independent part (\ie the vacuum energy)
from both
sides; we must in any case for consistency
since we have done so before (\cf \fact\ and below). This requires
separating from the above two-point function the (field independent) effective
self-energy:
\eqn\defGamhat{
 {\delta^2 \Gamma_\Lambda\over
\delta\phic_x\delta\phic_y} = \Sigma_{xy}
+{\hat\Gamma}_{xy}[\phic]\quad.}
Hence, using the fact that $\Sigma\equiv\Sigma(p;\Lambda)$ is diagonal in
momentum space
and \defK,
the subtracted version reads:
\eqn\smoothflow{
{\partial\over\partial\Lambda}\Gamma_\Lambda[\phic]
=-{1\over2}\tr\left\{
{K_\Lambda\over
(1+\Delta_>\Sigma)^2}.{\hat\Gamma}.
\left(1+[\ipropg\,+\Sigma]^{-1}\!.
{\hat\Gamma}\right)^{-1}\right\}\quad.}
This is the version of the 1PI flow equations with a smooth cutoff. We see that
there is no longer a term with an unintegrated $K_\Lambda$,
like the first term in \Pexpanded, responsible for the discontinuous
behaviour of $\Sv$ in the sharp cutoff limit. Indeed, since all
$\theta$-functions now appear
under the integral (trace), and they appear in such a way as to  have a
sensible limit, the R.H.S. is a smooth function
of $\Lambda$ in the limit $\epsilon=0$. The boundary conditions at
$\Lambda=\Lambda_0$ follow from \deltas\ and \GamtoS:
$\Gamma_{\Lambda_0}=S_{\Lambda_0}=$ bare action.

We may take the limit, using our little lemma \lemma\ and the explicit relation
\deltas\ and \defde. We specialize  to $\iprop\,(p)=p^2$ as before (\cf above
\twoP ).
This gives:
\eqn\sharpflow{{\partial\over\partial\Lambda}\Gamma_\Lambda[\phic]
=-{1\over2} \int\!{d^Dq\over (2\pi)^D}\ {\delta(q-\Lambda)\over
q^2+\Sigma(q;\Lambda)} \left[{\hat\Gamma}.
(1+G.{\hat\Gamma})^{-1}\right](\q,-\q)\quad,}
where now
\eqn\defG{G(p;\Lambda)\equiv{\theta(p-\Lambda)-\theta(p-\Lambda_0)\over
p^2+\Sigma(p;\Lambda)}\quad.}
The assumption made in using \lemma\ is that the term in \sharpflow\ enclosed
in square brackets, or equivalently ${\hat\Gamma}[\phic](\q,\p)$, is not
singular---is in fact continuous---at $\p=-\q$. This is true providing $\phi$
does not have a non-zero constant part  $\phie $ (\ie vacuum expectation value,
in momentum space a term of the form $\phi(\p)\sim (2\pi)^D\delta(\p)\phie$).
In the latter case we must
split off from ${\hat\Gamma}[\phic]$, the constant field part \ie instead of
\defGamhat\ above one writes:
\eqn\defGamvac{{\delta^2 \Gamma_\Lambda\over
\delta\phic_x\delta\phic_y} = \Sigma_{xy}(\phie )
+{\hat\Gamma}_{xy}[\phic]\quad.}
where $\Sigma_{xy}(\phie )$ is still diagonal in momentum space and defined by
\eqn\defSigvac{\Sigma_{xy}(\phie ) \equiv {\delta^2 \Gamma_\Lambda\over
\delta\phic_x\delta\phic_y}\Biggr|_{\phic=\phie }\quad .}
Working again from \thestart\ and dropping a field independent term we obtain
\eqn\expecflow{{\partial\Gamma_\Lambda[\phic]\over\partial\Lambda}
=-{1\over2} \int\!{d^Dq\over (2\pi)^D}\ \delta(q-\Lambda)\left\{{1\over
q^2+\Sigma} \left[{\hat\Gamma}. (1+G.{\hat\Gamma})^{-1}\right](\q,-\q)
+V\ln(q^2+\Sigma)\right\}.}
where now $\Sigma$ is given by \defSigvac\ \ie
$$\Sigma\equiv \Sigma(q;\phie,\Lambda)\ins11{and}
G(p;\Lambda)\equiv{\theta(p-\Lambda)-\theta(p-\Lambda_0)\over
p^2+\Sigma(p;\phie,\Lambda)}\quad,$$
and $V$ is the space-time volume.

Using $(1+G.{\hat\Gamma})^{-1}=1-G.{\hat\Gamma}+(G.{\hat\Gamma})^2-\cdots$ in
\sharpflow\ and expanding in $\phic$ one obtains the flow equations for the
vertices:
\eqn\flowexpanded{{\partial\over\partial\Lambda}
\Gamma(\p_1,\cdots,\p_n;\Lambda) =\int\!{d^Dq\over (2\pi)^D}\
{\delta(q-\Lambda)\over q^2+\Sigma(q;\Lambda)}
 E(\q,\p_1,\cdots,\p_n;\Lambda)}
where
\eqn\defE{\eqalign{& E(\q,\p_1,\cdots,\p_n;\Lambda)=
-{1\over2} \Gamma(\q,-\q,\p_1,\cdots,\p_n;\Lambda)\cr
&+\sum_{\{I_1,I_2\}}\Gamma(\q,-\q-\P_1,I_1;\Lambda)G(|\q+\P_1|;\Lambda)
\Gamma(\q-\P_2,-\q,I_2;\Lambda)\cr
&-\sum_{\{I_1,I_2\},I_3}\Gamma(\q,-\q-\P_1,I_1;\Lambda)
G(|\q+\P_1|;\Lambda)\times\cr
&\hskip 1cm\Gamma(\q+\P_1,-\q+\P_2,I_3;\Lambda) G(|\q-\P_2|;\Lambda)
\Gamma(\q-\P_2,-\q,I_2;\Lambda)\cr
 &\phantom{-\sum_{\{I_1,I_2\},I_3}\Gamma(\q,-\q-\P_1,I_1;\Lambda)
G(|\q+\P_1|;\Lambda)\times\ }\hskip 1cm
+\cdots\quad .}}
Here $\P_i=\sum_{\p_k\in I_i}\p_k$ and $\sum_{\{I_1,I_2\},I_3,\cdots,I_m}$
means a sum over all disjoint subsets $I_i\cap I_j=\emptyset$ ($\forall i,j$)
such that
$\bigcup_{i=1}^m I_i=\{\p_1,\cdots,\p_n\}$. The symmetrization $\{I_1,I_2\}$
means this pair is counted only once \ie $\{I_1,I_2\}\equiv\{I_2,I_1\}$, and is
a reflection of the Bose symmetry in the equation. Once again the equation is
best appreciated graphically---\cf fig.\fig\flowexpfig{ }.
\midinsert
\centerline{
\psfig{figure=exactrgfig2.ps,width=7in}}
\bigskip

\centerline{\vbox{{\bf Fig.2.} The flow equations for
the 1PI vertices in the sharp cutoff limit. Internal lines are full
propagators. The black dot now represents restriction to momentum $q=\Lambda$;
the other  propagators have an I.R. momentum cutoff $p>\Lambda$.}}
\endinsert
Evidently the expansion stops at the term where all 1PI vertices have their
minimum number of legs, \ie at the $n^{\rm th}$ term in general, or the
$(n/2)^{\rm th}$ term in the $\phi\leftrightarrow-\phi$ invariant theory. It is
often helpful
to rewrite \flowexpanded\ by performing the integral over $q$ and factoring out
the $D$-dimensional solid angle thus:
\eqn\angleflow{{\partial\over\partial\Lambda}\Gamma(\p_1,\cdots,\p_n;\Lambda)
={2\over\Gamma({D\over2})(4\pi)^{{D\over2}}}{\Lambda^{D-1}\over
\Lambda^2+\Sigma(\Lambda;\Lambda)}
 <E(\q,\p_1,\cdots,\p_n;\Lambda)>_{q=\Lambda}\ .}
In here $<\cdots>_{q=\Lambda}$ means an average over all directions for $\q$,
its length restricted to $q=\Lambda$.

Of course these equations can be derived by first expanding the smooth
cutoff equations \smoothflow\ and then taking the limit. Equations
\flowexpanded\defE\ are valid providing the  external momenta in the
propagators $G$  do not vanish (\ie $P_1,P_2\ne0$ \etc). If this happens then,
by \defG\ and \flowexpanded, we obtain $\theta(0)$'s which, as we have
already seen, are ambiguous. If nothing special happens at zero momentum we can
ignore this problem since they are points of measure zero in any calculation or
realistic
physics problem. On the other hand zero momentum is special if the field
acquires a vacuum expectation value; the correct equations either follow
from \expecflow, or expansion of \smoothflow\ and the setting of the relevant
momenta to zero before using the limit \lemma . Apart from the odd aside, we
will use \flowexpanded\defE\ from now on and ignore the more general case
\expecflow\ in this paper.

Equation \defG\ implies that a contribution
in \defE\ vanishes if an intermediate momentum, \eg $|\q+\P_1|$,  is larger
than $\Lambda_0$ and so has a profound effect on the solutions especially
for external momenta and/or $\Lambda$ close to (but less than) the cutoff.  If
one recalls however that $\Gamma_\Lambda$
may be regarded as just a smooth parametrization of the effective action
$S_\Lambda$ this result might seem a little puzzling: after all in the
Polchinski equation \Peqn\ contributions from momenta $p>\Lambda+\epsilon$ are
heavily suppressed, and in the limit $\epsilon=0$ one  na\"\i vely would expect
the equations to depend only on momenta $p<\Lambda$. As we have seen however,
although all the $\Sv$'s are restricted to $p_i<\Lambda$ in the limit
$\epsilon=0$, in the sense that any propagator connecting them is
$\Delta_<(p)=0$ for $p>\Lambda$ (\cf \ztozl\ and \zlresult),  the limit itself
is ambiguous. To resolve the limit as the cutoff becomes sharp it is necessary
to isolate the rapid change at the cutoff from the smooth parts. The latter
enter the limit in such a way that they need to be known at least as a series
expansion (about $p_i=\Lambda$) to all orders. In other words we need a
continuation of
the smooth parts to the region $p_i>\Lambda$. There are presumably many ways
to extend the smooth parts to $p_i>\Lambda$ but  {\it nota bene} however that
the resulting $\Sv$'s, in the regime $p_i<\Lambda$, {\sl will be invariant
under such choices}. If we require that these choices also correctly represent
1PI greens functions for $p_i>\Lambda$ then the number of possibilities is
severely restricted. The simplest possibility is to require that the
$\Gamma(\p_1,\cdots,\p_n;\Lambda)$'s are taken to be as analytic as
possible, and the minimal requirement of the  series expansion about
$p_i=\Lambda$ incorporated as
the maximally analytic extension into the region ${\rm Re}(p_i)>\Lambda$. This
is essentially what will be assumed in our approximation scheme. It is
clear that such a prescription involves dropping the $\Lambda_0$ terms in
\deltas\
and \defG, in other words taking the limit $\Lambda_0\to\infty$, since any
Taylor expansion about $p=\Lambda$ will not see the theta function
$\theta(p-\Lambda_0)$.

One can see that no U.V. singularities are introduced into the
 eqns.\smoothflow--\angleflow\  by the limit $\Lambda_0\to\infty$ since they
involve only an integration over $\q$ with fixed length $q=\Lambda$.
Of course U.V. divergences  will still arise in the solutions,  so we need
now to set the boundary conditions at some finite scale $\Lambda=\Lambda'_0$.
The most analytic choice will be obtained if the boundary values
$\Gamma(\p_1,\cdots,\p_n;\Lambda_0')$ are chosen to be  analytic functions of
momenta $p_i^\mu$, for localities sake we further choose the `bare'
$\Gamma_{\Lambda'_0}[\phic]$ to have a local derivative expansion. The relation
between the `bare' Wilsonian effective action $S_{\Lambda'_0}$ and this `bare
Wilsonian quantum effective action'
now involves the full tree expansion \GamtoS. One can expect though, by the
usual universality arguments, that these differences only alter the way the
cutoff effects are parametrized in any given theory (compensatable by a
reparametrization of the irrelevant couplings), while the universality classes
remain the same. However, the concept of a bare
action in a path integral as in \zorig\ appears now to be lost.

Actually nothing quite so drastic has happened: we know that we can reconstruct
the theory for $p<\Lambda'_0$ by using $S_{\Lambda_0'}$ as the bare
action and an overall momentum cutoff $\Lambda_0'$, and thus also a Legendre
effective action $\Gamma^{\Lambda_0'}_\Lambda$  if we use \eg \sharpflow\ and
\defG\ with $\Lambda_0\mapsto\Lambda_0'$ and the boundary condition
$\Gamma^{\Lambda_0'}_{\Lambda_0'}=S_{\Lambda_0'}$. We display explicitly with a
superscript the
fact that this new Legendre effective action has an overall cutoff
$\Lambda_0'$. From the boundary condition if nothing else, it evidently differs
from the $\Gamma^{\Lambda_0}_{\Lambda}\to\Gamma^{\infty}_{\Lambda}$ discussed
in the previous paragraph. By construction however it nevertheless
yields the same $S_\Lambda$ for all scales less than $\Lambda_0'$. We have thus
sketched how to recover a Legendre effective action computed from a bare action
at $\Lambda_0'$ by a reparametrization of the Legendre
effective action
$\Gamma^{\Lambda_0}_{\Lambda}\mapsto\Gamma^{\Lambda_0'}_\Lambda$, while leaving
the physics at scales lower than $\Lambda_0'$ untouched. This is thus an
invariance of the form stressed above.

We now fill in the details. As already indicated in the introduction the
reparamet\-rization is itself just a tree diagram expansion. Subtracting the
field independent term $\half\tr\{(d\ipropg\,\big/d\Lambda).\Delta_>\}$
from \thestart\ we have that
\eqn\gloop{
{\partial\Gamma_\Lambda^{\Lambda_0}[\phi]\over\partial\Lambda}\equiv
{1\over2}\tr\left\{{d\Delta_\Lambda^{\Lambda_0}\over d\Lambda}.\left(
\left[{\delta^2\Gamma_\Lambda^{\Lambda_0}\over\delta\phi\delta\phi}\right]^{-1}
+\Delta^{\Lambda_0}_\Lambda\right)^{-1}\right\}\quad.}
In here we have explicitly displayed the U.V. and I.R. cutoffs on
$\Delta_>=\Delta^{\Lambda_0}_\Lambda$ with
\eqn\twocuts{
\Delta^{\Lambda_0}_\Lambda \equiv \left[ \te p - \theta_\epsilon(p,\Lambda_0)
\right]\ \Delta(p)\quad,}
as given in \deltas . We have dropped the superscript ${}^c$ on the classical
fields to unclutter the notation. Define the required tree diagram relations
through the generalised Legendre transform:
\def\glzp{\Gamma_\Lambda^{\Lambda_0'}}
\def\glz{\Gamma_\Lambda^{\Lambda_0}}
\def\dlzplz{\Delta_{\Lambda_0'}^{\Lambda_0}}
\def\dllzp{\Delta_{\Lambda}^{\Lambda_0'}}
\def\dllz{\Delta_{\Lambda}^{\Lambda_0}}
\eqn\changetop{
\glzp[\phi']=\glz[\phi]+{1\over2}(\phi-\phi').\left(\dlzplz\right)^{-1}\!\!
.(\phi-\phi')\quad.}
$\dlzplz$ is defined as in \twocuts, and here as elsewhere
$\Lambda_0'<\Lambda_0$. The tree expansion for $\glzp$ follows by solving for
$\phi$ iteratively via
\eqn\dphi{
{\delta\glz[\phi]\over\delta\phi}+\left(\dlzplz\right)^{-1}\!.(\phi-\phi')
=0\quad,}
as
\eqn\treeinv{
\phi=\phi'-\left[(\dlzplz)^{-1}+\Sigma^{\Lambda_0}_\Lambda\right]^{-1}
\!.{\delta\glz[\phi']\over \delta\phi'}+\cdots}
and substituting back into the R.H.S. of \changetop . Here
$\Sigma^{\Lambda_0}_\Lambda$ is the self-energy derived from the $2^{\rm nd}$
derivative of $\glz$.
(Note the close similarity of \changetop\ and \SG . The analogous expansion to
\treeinv\ gives the relation between $S_\Lambda$ and $\Gamma_\Lambda$ in terms
of full propagators.)

It is instructive to perform this for low orders and see diagrammatically how
the tree expansion simply shifts the overall cutoff in the equations
\flowexpanded \defE \defG . Here we will confirm it in full, analytically. From
\changetop\
we also have
\eqn\dphip{
{\delta\glzp[\phi']\over \delta\phi'}+\left(\dlzplz\right)^{-1}\!.(\phi-\phi')
=0\quad,}
which differentiating again with respect to $\phi'$ gives
$${\delta\phi_x\over\delta\phi'_y}
\left({\delta^2\glzp\over\delta\phi'\delta\phi'}\right)^{-1}_{yz}=
\left({\delta^2\glzp\over\delta\phi'\delta\phi'}\right)^{-1}_{xz}-
\left(\dlzplz\right)_{xz}\quad.$$
On the other hand, comparing \dphi\ and \dphip\ we have
$\delta\glzp\big/\delta\phi'=\delta\glz\big/\delta\phi$
which on differentiating again with respect to $\phi'$ gives
$${\delta\phi_x\over\delta\phi'_y}\left({\delta^2\glzp\over
\delta\phi'\delta\phi'}\right)^{-1}_{yz}=
\left({\delta^2\glz\over\delta\phi\delta\phi}\right)^{-1}_{xz}\quad.$$
Combining the last two displayed equations, and using $\dllz=\dlzplz
+\dllzp$ as follows from \twocuts, we have
\eqn\thecrux{\left({\delta^2\glz\over\delta\phi\delta\phi}\right)^{-1}
+\dllz = \left({\delta^2\glzp\over \delta\phi'\delta\phi'}\right)^{-1}
+\dllzp\quad.}
Together with the identity $d\dllz\big/d\Lambda =d\dllzp\big/d\Lambda$,
which trivially follows from \twocuts, and the identity
$${\partial\over\partial\Lambda}\Biggr|_\phi\glz[\phi]=
{\partial\over\partial\Lambda}\Biggr|_{\phi'}\glzp[\phi']$$
which follows in the usual way from \changetop\ and \dphip, we see that
\thecrux\ implies that the flow equations \gloop\ are unchanged by the
reparametrization \changetop\ \ie if $\glz$ satisfies the flow equations then
so does $\glzp$ (and {\it vice versa}). It only remains to show that physics
with momentum $p<\Lambda_0'$ is unchanged. But this follows because the new
flow
equations, \gloop\ with $\Lambda_0\mapsto\Lambda_0'$, have the correct boundary
condition $\Gamma_{\Lambda_0'}^{\Lambda_0'}[\phi']=S_{\Lambda_0'}[\phi']$.
To confirm this last point set $\phic\equiv\phi$, $\Phi=\phi'$ and
$\Lambda=\Lambda_0'$ in \SG;
recall that in \SG\ everything has overall cutoff $\Lambda_0$ and compare with
\changetop . The result follows immediately.

\section{Approximation Methods.}
In this section we consider methods of approximating the sharp cutoff
flow equations \flowexpanded--\angleflow\ where we drop the overall momentum
cutoff as discussed at the end of the previous section. We start however by
making some general comments on the non-local effects
induced by taking the sharp cutoff limit. We then turn to the flow equations
with smooth cutoff: these do not suffer from non-localities, but
have the serious calculational problems  already mentioned in the introduction.
After considering the straightforward numerical approach, we look at the
most promising method, an expansion in external momenta about $p_i=0$ in the
sharp cutoff limit. This is briefly compared to two alternative expansions.

The non-localities correspond to non-analytic dependence on the momenta at
$p^\mu_i=0$.
We have already seen an example of this in the analysis of
\defGamvac--\expecflow\ where the constant part of $\phi$ had to be treated
separately. This is connected to the fact that the solutions to \flowexpanded,
equivalently \angleflow,
generally do not have well defined limits as any $\p_i\to{\bf0}$, rather they
remain
functions of the relative angles between the $\p_i$ even in the limit. One can
see this by inspection of the third term of \defE, however first let us look
at the second  term. In the greens function $G(|\q+\P_1|;\Lambda)$ there is a
term $\theta(|\q+\P_1|^2-\Lambda^2)\approx\theta(\q.\P_1)$ for
$P_1<\!<\Lambda$. Expanding the $\Gamma$'s in the second term of \defE\ in this
limit we see that there are terms of the form
$<\q.\P_1\theta(\q.\P_1)>_{q=\Lambda}\sim \Lambda|\P_1|$, so that
\def\Gv{\Gamma(\p_1,\cdots,\p_n;\Lambda)}
$\Gv$ has an expansion in lengths $|\p|$ and not $\p^2$ as would be required by
locality/analyticity. In the third term one also gets contributions of the form
$<\theta(\P_1.\q)\theta(\P_2.\q)>_{q=\Lambda}$ when $P_1,P_2<\!<\Lambda$ which
depend only on the angle between $\P_1$ and $\P_2$ and not on their
lengths at all.

Disturbing as these non-localities might at first appear they are all
understandable
in the intuitive sense already covered in the introduction. Indeed one can see
all these effects  already in perturbation theory \ie in  Feynman diagrams, if
one provides each propagator with a sharp I.R. cutoff (as in \defG). We have
searched for any deep problems with using a sharp cutoff, this is a primary
reason for analysing the simple models in the next section, but we have not
uncovered any. One obvious constraint is that the resulting physics in the
limit $\Lambda\to0$ does not suffer from these non-localities. In other words
that the non-localities present in the intermediate  effective action
$\Gamma_\Lambda$ are cancelled by those in the remaining physics with momenta
constrained to have  $p<\Lambda$. But this is
guaranteed if we  keep the bare $\Gamma_{\Lambda_0}[\phi]$ local.\foot{We now
drop the primes introduced at the end of the last section.}\ (Presumably it is
possible to be more subtle,  noting that since the typical effective action
$\Gamma_\Lambda$ is non-local the bare effective action can be too, providing
the non-localities have some closely prescribed form). It seems possible that
non-local terms in $\Gamma_\Lambda$ could turn out to be relevant---in other
words have to be added as counterterms to the bare lagrangian, making it
non-local, in order to renormalize the low energy theory. It is clear that this
does not happen in perturbation theory from the interpretation in terms of
Feynman diagrams above; it is also clear that it cannot happen \nonp ly in any
truncation, by thinking about the smooth case first and then taking the limit
$\epsilon\to0$
(because the divergent parts of the bare couplings will not depend on
$\epsilon$ in this limit).
But it is  allowed  by na\"\i ve power counting. If this would happen  as the
result of some approximation in the momentum dependence (\cf the expansions
later), it would  be a disaster: the low energy physics would likely be
non-local also,
while new and unphysical parameters would have been added to  the bare
lagrangian. As we
will see in the next section this doesn't happen, at least in the situations we
considered.

It is intuitively clear that taking the sharp cutoff limit ought to simplify
matters, but given the non-local behaviour sketched above it is worthwhile
taking a careful look at approximating the smooth cutoff equations \smoothflow.
This can be done by making  a momentum expansion around $\p_i={\bf0}$ and
truncating at some order. Recall the general form of $\theta_\epsilon$ as
described below \deltas. From there it is clear we must have
$\epsilon<\Lambda$, otherwise there will be no effective cutoff on low momenta.
Similar $\theta$ terms as in the sharp cutoff case
are of course encountered but now they appear as $\te{|\q+\p|}$ where $\p=$ \eg
$\P_1,\P_2$ \etc  Expanding these about $\p={\bf0}$ gives
\eqn\smooTaylor{
\eqalign{
\te{|\q+\p|}=&\te q +{\q.\p\over q}{\partial\over\partial q}\te q\cr
&+{1\over2}\left\{ {p^2q^2-(\q.\p)^2\over q^3}{\partial\over\partial q}\te q
+{(\q.\p)^2\over q^2}{\partial^2\over\partial q^2}\te q \right\}
+\cdots\quad.\cr}
}
We  see that because $\theta_\epsilon$ is a smooth function a true Taylor
expansion  in the $p_i^\mu$ about $p_i^\mu=0$ now exists for the $\Gv$,
equivalently $\Gamma_\Lambda[\phi]$ has a local derivative expansion, with
radius of convergence
$p\approx\epsilon$. This radius follows from \smooTaylor\
 since we have by simple dimensional arguments $(\partial\big/\partial q)^n
\theta_\epsilon(q,\Lambda)\approx \half\epsilon^{-n}$ for $q\approx\Lambda$,
and since $\epsilon<\Lambda$. Substituting the expansion
of $\Gamma_\Lambda$  into \smoothflow\ however involves integrating the
expansion over a momentum shell with radius $\Lambda$, thus for convergence of
the expansion we need $\Lambda<\epsilon$. We see we need both
$\epsilon<\Lambda$ and $\epsilon>\Lambda$. We can relax each of these to
equality but then we have
only a weak suppression of low momenta and an expansion at its radius of
convergence where it will converge only very weakly if at all.
\def\erf{{\rm erf}}

To justify the estimated  radius of convergence we look at a few examples of
$\theta_\epsilon$. If $\theta_\epsilon$ is an exact cutoff on either side \ie
$\te q=0$ for all $q>\Lambda+\epsilon$ and/or for all $q<\Lambda-\epsilon$ then
the radius of convergence of the expansion \smooTaylor\ with $q\approx\Lambda$
is at most $\epsilon$ from elementary complex analysis. An example of an
approximate cutoff is
$$\theta_\epsilon(q,\Lambda)=
\half\left(1+\tanh[{\textstyle{3\over2}}
(q-\Lambda)/\epsilon]\right)\quad.$$
A plot of $\tanh(3x/2\epsilon)$ will convince the reader that this has width
$\approx2\epsilon$ as required ($\tanh(3/2)=0.9$). On the other hand the radius
of convergence of \smooTaylor\ is $p=\pi\epsilon/3\approx\epsilon$ as given by
the first pole of $\tanh(z)$ on the imaginary $z$-axis. Actually it is
technically possible to have infinite radius of convergence, but not in any
practical sense since for $\epsilon\le\Lambda$ the series only starts to
converge at high order.
For example  write
\eqn\analcutoff{\te q
=\half\left(1+\erf\!\left[a(\q^2-\Lambda^2)
\big/2\Lambda\epsilon\right]\right)\quad,}
where erf is the error function and $a$ is to be chosen so that \analcutoff\
has width $2\epsilon$.
Clearly the convergence is best if we take the maximum width
$\epsilon=\Lambda$. We require $a\approx2.5$ to get effective suppression of
low momenta ($\theta_\Lambda(0,\Lambda)=0.04$), \ie to get the correct width in
the $q<\Lambda$ regime. But the fact that $q$ appears squared in \analcutoff\
makes the $q>\Lambda$ behaviour very sharp, and this has serious effects on the
effective convergence of the series. Indeed estimating the convergence in the
flow equations by expanding \analcutoff\ about $\Lambda$ as $q=\Lambda+\Delta
q$, and then setting $\Delta q=\pm\Lambda$, one finds that 10\% accuracy
requires expansion to 84$^{\rm th}$ order!

The second major problem with the smooth cutoff case is as follows.
Substituting \smooTaylor\  (and an expansion of $\Gamma_\Lambda$) into
\smoothflow, one sees that the coefficients of the momentum expansion for
$\Gamma_\Lambda$ are determined by integrals over $q$ of
$\delta_\epsilon(q,\Lambda)$ times terms containing $(\partial\big/\partial
q)^n \theta_\epsilon(q,\Lambda)$. Unlike the terms in \lemma, these
coefficients have no
universal behaviour for $\epsilon<\!<\Lambda$, but depend sensitively on the
shape of the cutoff.\foot{Of course they  diverge as $\epsilon\to0$, but we
mean also that the coefficient of the divergence has no limit.}\
While this is reasonable, given that these higher derivative corrections are
`mixing' with the $\theta_\epsilon(q,\Lambda)^{-1}$ in the kinetic term---the
latter corresponding to  a higher derivative regularisation with an infinite
number of derivatives, the lack of universality makes their interpretation
difficult. Furthermore, if we take $\epsilon\sim\Lambda$  the sensitivity to
the form of the
cutoff  is  physical since it reflects how much of the low momentum modes
$q<\Lambda$ are integrated out.

There are a few less major problems with the smooth cutoff case also: the above
integrals can be cast as one-dimensional integrals over $q$ but they still have
to be done numerically, and since the integrals depend on $\Sigma(0;\Lambda)$,
\smoothflow\ becomes a set of integro-differential equations. Also the
imposition of a wide smooth cutoff makes matching to other methods---\eg
numerical, perturbative \etc---problematical. Given all these problems, it
seems clear that local approximations to the smooth flow equations can not be
extended in any substantive way beyond the uncontrolled approximation for the
effective potential (covered in the introduction) where it differs
inessentially from the sharp cutoff limit.

Let us note here that attempts to solve either the smooth or sharp cutoff flow
equations directly by numerical means will also encounter severe problems. We
would need to discretize over the arguments of the greens functions, replacing
$\partial\big/\partial\Lambda$ by a finite difference and the integration over
spherical shells by the appropriate sums.  At least na\"\i vely, we must work
with the values of the  $\Gv$ over all combinations of discretized values for
the $p^\mu_i$ $i=1,\cdots,n$ $\mu=1,\cdots,4$ in the range $0<p_i<\Lambda_0$.
Even allowing for rotational invariance, permutation symmetry, and momentum
conservation, this quickly gets out of hand as $n$ increases. The problem is
not so much the storage of so many numbers but the fact that they all have to
be updated as $\Lambda\mapsto \Lambda-\delta\Lambda$. It is easy to convince
oneself that the computational difficulty is greater than that of lattice gauge
theory even if we consider only $n\le6$. There seem to be many possibilities
for improving this
na\"\i ve scenario but the crucial observation is that much of the integration
domain in $(\p_1,\cdots,\p_n;\Lambda)$ space must be boring: this suggests that
we should try to handle these regions semi-analytically, \eg by the momentum
expansion which follows, leaving numerical analysis for (hopefully) small
ranges where semi-analytic methods may fail. Actually, semi-analytic methods
could be pushed very far before their numerical complexity rivalled that of
straightforward numerical approaches.

We turn at last to the most promising approach. We work with the sharp cutoff
equations \defE\angleflow. Writing all mass scales in units of $\Lambda$
utilises the scaling symmetry in the equations. Thus we replace $\Lambda$ with
$t=\ln(\Lambda_0/\Lambda)$ (initial b.c's are set at $t=0$), and write
%\eqn\defgam{
$\Gamma(\Lambda\p_1,\cdots,\Lambda\p_n;\Lambda)\equiv
\Lambda^{d_n}\gamma(\p_1,\cdots,\p_n;t)$
%\quad,}
 where $d_n=D+(1-D/2)n$ is the engineering (or canonical) dimension of $\Gv$.
Similarly we write $\Sigma(\Lambda p;\Lambda)\equiv \Lambda^2 \sigma(p;t)$.
Finally we replace $\p_i$ by $\rho\p_i$ where $\rho$ is the `momentum scale'
and is our expansion parameter. (It may be set to one at the end). This gives
\eqn\scaledflow{\eqalign{\left(
{\partial\over\partial t}+\rho{\partial\over\partial\rho}
-d_n\right)&\gamma(\rho\p_1,\cdots,\rho\p_n;t)=\cr
&-{2\over(4\pi)^{{D\over2}}\Gamma({D\over2})} {1\over1+\sigma(1;t)}
<E(\q,\rho\p_1,\cdots,\rho\p_n;t)>_{q=1}\ .\cr}
}
$E$ is still given by formula \defE, (with $\Gamma$'s replaced by $\gamma$'s)
see also fig.\flowexpfig, while $G$ is now given by
\eqn\defGtrue{
G(p;t)\equiv{\theta(p-1)\over p^2+\sigma(p;t)}\quad.}
Now we expand in small momentum scale as follows
\eqn\expgam{
\gamma(\rho\p_1,\cdots,\rho\p_n;t)=\sum_{m=0}^\infty \rho^m
\gamma_m(\p_1,\cdots,\p_n;t)}
so the $\gamma_m$  scale homogeneously:
\eqn\defgm{
\gamma_m(\rho\p_1,\cdots,\rho\p_n;t)=\rho^m \gamma_m(\p_1,\cdots,\p_n;t)\quad.}
This expansion incorporates the non-local effects covered at the beginning of
this section since the integer $m$ is not restricted to being even and \eg
$\gamma_0$ is not necessarily constant but is in general a function of the
angles between the $\p_i$. The R.H.S. of \scaledflow\ may be evaluated as
follows. Let $x=\p.\q/p=\cos\vartheta$, where $\vartheta$ is the angle between
$\q$ and $\p$,  then expansion of the I.R. cutoff is achieved as
\eqn\expIR{
\theta(|\rho\p+\q|-1)=\theta(x+\rho
p/2)=\theta(x)+\sum_{n=1}^\infty{\rho^n\over
n!}\,({p\over2})^n\,\delta^{(n-1)}(x)\quad.}
Terms of the form $\gamma_m(\q+\rho\P_1,-\q+\rho\P_2,\cdots)$ (the most
general) have re-expansions in $\rho$.  The angular average may be performed,
for example  in four dimensions,  writing $\q=(\cos\vartheta,\sin\!\vartheta\,
{\underline n})$ where ${\underline n}$ is a unit 3-vector:
\eqn\average{
<\cdots>_{q=1}={2\over\pi}\int^\pi_0\!\!d\vartheta\,\sin^2\!\vartheta\,
<\cdots>_{{\underline n}}\ \
={2\over\pi}\int^1_{-1}\!\!dx\sqrt{1-x^2}<\cdots>_{{\underline n}}
\quad.}
The R.H.S.  has an average over azimuthal directions.
The approximation is achieved by working to some maximum order in $\rho$ for
each $n$-point function, and to some maximum $n$ \eg by setting higher
$n$-point functions arbitrarily to zero. Since \scaledflow\ is a set of
first-order differential equations with boundary conditions fully determined by
the $\gamma(\p_1,\cdots,\p_n;0)$ of the `bare' action, the solutions to these
truncations are unique.
In the cases we discuss in the next section the solutions for $\gamma_m$ factor
into simple functions of the momenta times $t$ dependent coefficients.

This approximation can be expected to work well if  rapid variations with
respect to  momenta happen not at all, or at a scale $p$ such that
$p/\Lambda<\!<1$. In addition if the $\gamma$'s  scale approximately as $\sim
\Lambda^{-d_n}$  (\ie $\Gamma$'s $\sim$ constant) below some mass scale $M$
then the expansion is really in $p/M$. Mathematically one expects the radius of
convergence to be the maximum one determined by the first singularity in the
complex $\rho$ plane.

As discussed already in the introduction the crucial simplification in this
approximation scheme is the expansion of the I.R. cutoff \expIR. There are a
couple of other realistic  systematic expansions, but they are probably of
limited use. One is to
reintroduce an overall cutoff $\Lambda_0$: After scaling out $\Lambda$,
everything may be expanded  in $t=\ln(\Lambda_0/\Lambda)$. Thus the cutoff
terms (\cf \defG) become:
\eqn\simply{\theta(p-1)-\theta(p-\e{t})=
t\delta(p-1)+{t^2\over2}[\delta(p-1)-\delta'\!(p-1)]+\cdots}
and $\gamma(\p_1,\cdots,\p_n;t)=\sum_{m=0}^\infty t^m
\gamma_m(\p_1,\cdots,\p_n)$. The $\Gamma_0$'s are the bare vertices. Evidently
this scheme  only works  for $\Lambda$ not much different from $\Lambda_0$. If
$\Lambda_0$ is the true cutoff such an expansion will only see cutoff effects,
so it is only useful if $\Lambda_0$ is  an effective cutoff, \eg introduced via
the reparametrization discussed at the end of sect.3. The expansion \simply\
introduces via the angular average (where $p$ is replaced by $|\P_k+\q|$)
powers of $1/P_k$. This expansion does not therefore converge for small momenta
and does not match easily on to the above momentum expansion.

Another possibility is (again) to drop $\Lambda_0$,  not scaling out $\Lambda$
but working with \defE\ and \angleflow, and expanding in  $\Lambda<\!< p_i$.
This implies also Taylor expanding in $q^\mu$. In this regime we can set all
$\theta$ terms to 1. The angular average is easily computed by the identities
$<q^\mu q^\nu>_{q=\Lambda}=\delta^{\mu\nu}\Lambda^2/D$ and their
generalisations. This expansion is easy to perform and works well in the,
albeit limited, regime $\Lambda<\!<p_i$.

\section{Examples.}
In this section we investigate two simple models of four dimensional \pf\
to test the approximation scheme \scaledflow-\average. The first is the
analogue of the ladder (or cactus) approximation used in \DS: it allows a
comparison and checks on the effect of a sharp cutoff. The second  model
incorporates the first \nonp\ correction to the coupling and provides some
tests of the momentum expansion. We also check that no non-local terms become
relevant. An overview of the results has already been given in the
introduction.

To obtain the cactus approximation to the flow equations \scaledflow\ we
truncate to the $n=2$ equation:
\eqn\twoflow{\left(
{\partial\over\partial t}+p{\partial\over\partial p}
-2\right)\sigma(p;t)=
{1\over(4\pi)^2} {1\over1+\sigma(1;t)} <\gamma(\q,-\q,\p,-\p;t) >_{q=1}\ .
}
Now we substitute the ansatz $\gamma(\p_1,\cdots,\p_4;t)=\lambda$
which corresponds in particular to ignoring the running of the coupling. It
also implies in this approximation that $\sigma\equiv\sigma(;t)$ has no
momentum dependence. Thus, writing $\alpha=\lambda/(4\pi)^2$:
\eqn\cactusflow{
{\partial\sigma\over\partial t}=2\sigma+{\alpha\over 1+\sigma}\quad.}
 While we ignore the running of $\alpha$ in this equation, we  keep in mind
that as we flow to low energies (\ie increasing $t$) $\alpha$ shrinks.\foot{as
follows \eg from the 1-loop $\beta$-function, but can also be seen
qualitatively directly from the $n=4$ equation. See later.}\ The shrinking will
continue to $\alpha=0$ at $\Lambda=0$ (\ie $t=\infty$), the gaussian fixed
point, unless $\sigma$ becomes bigger than one below some scale $\Lambda=M$. In
the latter case we expect that $\alpha$ freezes out at a value
$\approx\alpha(M)$ for scales $\Lambda<M$ (while $\Sigma(;0)\sim M^2$). Keeping
in mind this qualitative behaviour will ensure that we interpret the results of
\cactusflow\ correctly.

Consider first the case where the physical mass-squared $\Sigma(;0)>0$. In this
case the solution to \cactusflow\ is easily figured out:
\eqn\flowsoln{\ln\!\left({\Sigma(;0)\over\Lambda_0^2}\right)
={1\over2}\left\{\ln\!\left(\sigma_0(1+\sigma_0)
+{\alpha\over2}\right)+{1\over\sqrt{1-2\alpha}}
\ln\!\left({1-\sqrt{1-2\alpha}+2\sigma_0\over
1+\sqrt{1-2\alpha}+2\sigma_0}\right)\right\}\ ,}
where $\sigma_0=\sigma(;0)=m^2_0/\Lambda_0^2$.

Equation \cactusflow\ can be recast as an integral equation in a suggestive
way. It is easiest to go back  and integrate both sides of \flowexpanded\
between $\Lambda$ and $\Lambda=\Lambda_0$, again with $n=2$ and
$\Gamma(\p_1,\cdots,\p_4;\Lambda)=\lambda$:
\eqn\cactus{\Sigma(;\Lambda)=m_0^2+{\lambda\over2}
\int^{\Lambda_0}_\Lambda\!\!{d^Dq\over (2\pi)^D}\ {1\over
q^2+\Sigma(;q)}\quad.}
The range of the integration is $\Lambda<q<\Lambda_0$ as indicated.
For $\Lambda=0$ this is the cactus approximation  to the \DS, except for one
alteration: the self-energy inside the integral is not $\Sigma(;0)$ but is
evaluated at scale $\Lambda=q$. One might expect that such an alteration gives
a better approximation. The cactus approximation to the \DS\ gives a constant
$\Sigma$, determined implicitly by:
\eqn\DSsoln{
\Sigma=m_0^2+{\alpha\over2}
\left[\Lambda_0^2-\Sigma\,\ln\!\left(1+{\Lambda_0^2\over\Sigma}
\right)\right]\quad.}
To get some idea by how much the two approximations \flowsoln\ and \DSsoln\
differ, we compare them at strong and weak coupling. At weak coupling they
agree to order $\alpha$ (the result obtained by putting $\Sigma=m_0^2$ in the
R.H.S. of \DSsoln). This is simply because they both correctly incorporate the
one-loop correction. More non-trivial is agreement up to a factor of $\sqrt{2}$
at strong coupling. For the
flow equations:
$${\Sigma\over\Lambda_0^2}=\sqrt{\alpha\over2}+{\pi\over4}+
O\left({1\over\sqrt{\alpha}}\right)\quad.$$
For the \DS:
$${\Sigma\over\Lambda_0^2}={\sqrt{\alpha}\over2}+{m_0^2\over2\Lambda_0^2}
-{1\over3}+O\left({1\over\sqrt{\alpha}}\right)\quad.$$
Since the two cactus approximations  agree at small $\alpha$ and differ by 40\%
as $\alpha\to\infty$ they can be expected to be in rough agreement for all
$\alpha>0$ and   for all {\sl reasonable}  values of $m_0^2/\Lambda_0^2$. They
differ significantly when one starts to try and tune $\Sigma<\!<\Lambda_0^2$.
This is covered later.

To better appreciate the origin of the difference between the two
approximations, iterate the equation \cactus\ to get the perturbative
expansion. The result is given in terms of Feynman graphs in
fig.\fig\cactusfig{ }.
\midinsert
\centerline{
\psfig{figure=exactrgfig3.ps,width=6in}}
\bigskip

\centerline{\vbox{{\bf Fig.3.} Diagrams contributing to the cactus
approximation to the
flow equations. The loops have the I.R. cutoffs shown in their centres.}}
\endinsert
It is the same graphical expansion as that of the cactus approximation in \DS\
except that part of the momentum integrations are missing as a result of I.R.
cutoffs. For $\Lambda=0$ the disagreement shows first at two loops. The
remainder of this diagram is provided by adding the one-loop correction to the
4-point vertex (in the ansatz above \cactus) as indicated by the dotted box.
(This gives the same diagram except that now the lower loop is I.R. cutoff by
the higher loops momentum. Exchanging the order of integration here gives a
contribution in which the upper loop is U.V. cutoff by the lower loops
momentum, so the addition of the two contributions gives the two-loop diagram
with momenta integrated over the whole range.) Of course if we incorporate the
one-loop correction to the 4-point vertex then, by the symmetry of the flow
equations we also generate the other two-loop self-energy graphs, as well as
parts of many others. In contrast the \DS\ organise the graphical expansion
topologically; the term neglected in the Dyson-Schwinger equation for the
self-energy generates all graphs in which the two external legs are joined to
separate vertices.

It is clear from \flowsoln\ that the cactus approximation to the flow equation
has a different behaviour for $\alpha<\half$ compared to $\alpha>\half$. The
latter can be obtained from \flowsoln\ by analytic continuation, noting the
reality of the solution, but the qualitative nature of the solutions is best
appreciated by studying the R.H.S. of the flow equation \cactusflow\ as a
function of $\sigma$. This is sketched in fig.\fig\fixedpointfig{ }.
\midinsert
\centerline{
\psfig{figure=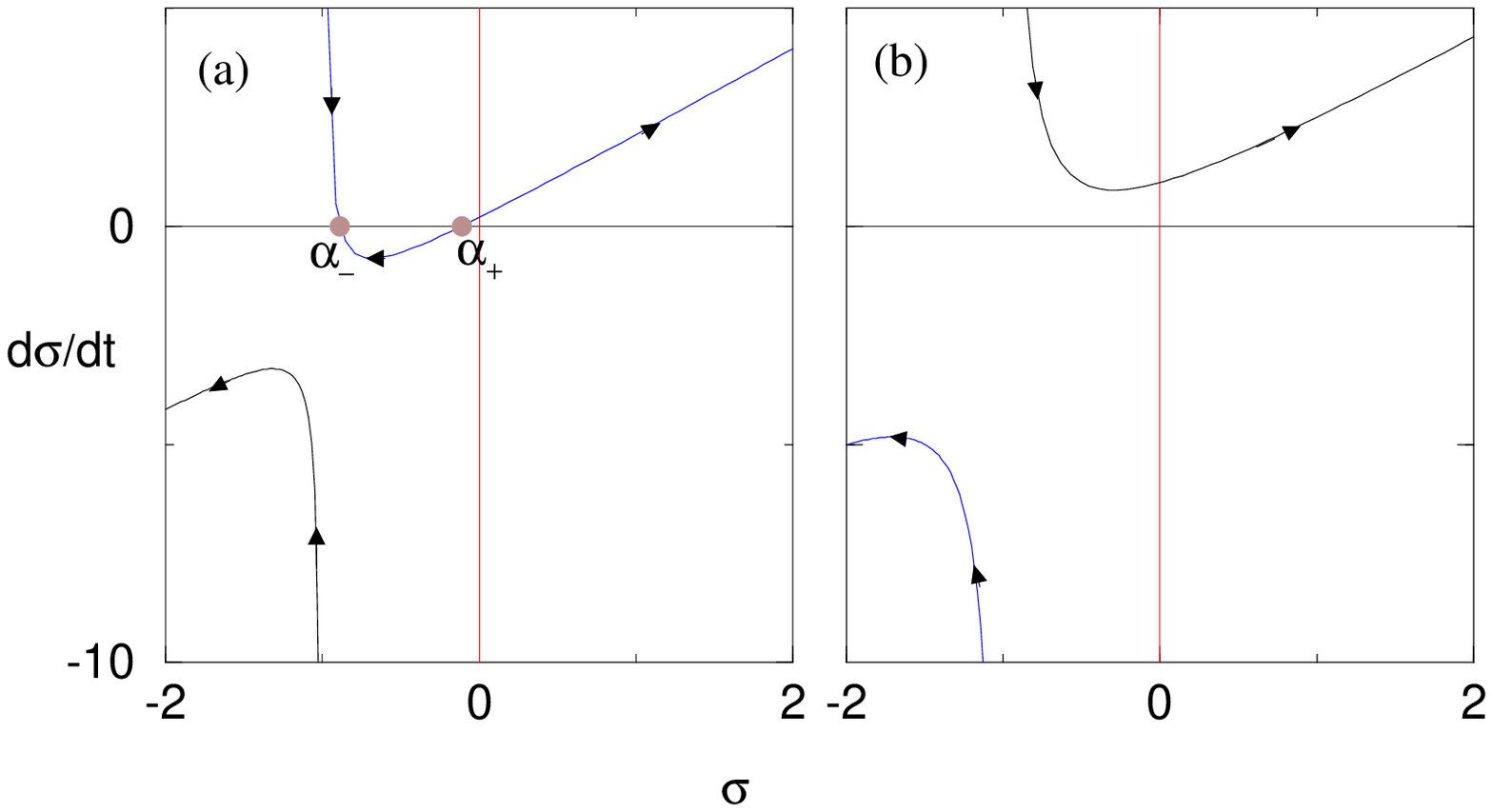,width=6in}}
\vskip -3.5in
\centerline{\vbox{{\bf Fig.4.} ${\partial\sigma/\partial t}$ as a function of
$\sigma$ in the cactus
approximation for (a) $\alpha<\half$, (b) $\alpha>\half$. Direction of flow for
increasing $t$ is indicated by arrows.
}}
\endinsert
 In this figure $\alpha_\pm$ are the
zeroes of the R.H.S: $\sigma=\alpha_\pm\equiv-\half\pm\half\sqrt{1-2\alpha}$.
For $\alpha\le\half$ it is clear  that if we require $\Sigma(;0)>0$ we must
have $\sigma_0>\alpha_+$. In this case as $t\to\infty$ we have
$\sigma\to\infty$ such that ${\partial\sigma/\partial t}\to2\sigma$ \ie
$\Sigma(;\Lambda)\to$ a positive finite constant, as $\Lambda\to0$. The
constant is  of order $\Lambda_0^2$ for generic $\sigma_0$ (\eg by dimensional
arguments). Since $\alpha_+$ is an I.R. unstable fixed point we can tune to
$\Sigma(;0)<\!<\Lambda_0^2$ by taking $m^2_0\to\alpha_+\Lambda_0^2$; indeed if
$m^2_0=\alpha_+\Lambda_0^2$ then $\Sigma(;0)=0$. This behaviour is similar to
that of the Dyson-Schwinger cactus approximation \DSsoln\ where, for all
$\alpha$, $\Sigma$ may be tuned to zero by $m^2_0\to-\half\alpha\Lambda_0^2$.

For $\alpha>\half$ however, there are no finite fixed points in
fig.\fixedpointfig. Now
 $\Sigma(;0)\sim\Lambda_0^2$ or larger for all choices of $m_0^2$. ( The phase
given by $m_0^2<-\Lambda_0^2$ is covered  below). This behaviour is not
modelled by the cactus approximation to the \DS. Is it reasonable? Keeping in
mind the comments below \cactusflow\ we see that we should interpret
$\alpha=\half$ as an effective maximum renormalized coupling $\lambda_{\rm
max}\approx8\pi^2=79.0$, above which there is no continuum limit because the
renormalized mass is of order the cutoff. This is therefore already a
reflection of triviality of four dimensional \pf. The accurate analysis
of L\"uscher and Weisz\adel\ gives $\lambda_{\rm max}=78\pm3$ (!). (See table 3
of that paper. Note $m_R^2\equiv\Sigma(;0)/\Lambda_0^2$ and
$g_R\equiv\lambda$).
Of course the {\sl exact} agreement is an accident since the
cactus expansion is only a crude approximation, and uses momentum cutoff, not
lattice cutoff as  in ref.\adel.

Now consider the  phases where $\sigma(;t)$ flows to negative values. For
$\alpha<\half$ we have, by fig.\fixedpointfig, an I.R. stable fixed point
$\sigma(;t)\to\alpha_-$ as $t\to\infty$, for all bare masses in the range
$-1<\sigma_0<\alpha_+$. Thus the physical mass $\Sigma(;0)=0$ for all bare
masses in this range. If now we again recall the qualitative evolution of
$\alpha$  given above \cactusflow\ we find reason to doubt this continuum
limit. Indeed we expect that $\alpha\to 0$ as $\Lambda\to 0$ in this regime,
but this would cause the fixed point $\alpha_-\to-1$, the tachyonic singularity
in fig.\fixedpointfig. From our general comments on convergence below \average,
and explicitly from the factors \defGtrue\ in the expansion \defE, one can see
that our expansion, of which the cactus approximation is the lowest order,
would then break down. To understand what really happens for this range of bare
parameters it is necessary to go beyond this simplest approximation. For
example it is clear that the problem is avoided  by first  shifting the field
$\phi$ \eg to the minimum of the potential, but it may be that the problem is
resolved after a reparametrization or as a result of unexpected behaviour in
the flow of $\lambda$.

Finally, for all $\alpha>0$ we have that $m_0^2<-\Lambda_0^2$ yields a finite
`physical' mass $\Sigma(;0)<-\Lambda_0^2$. Now however, expanding about the
unstable symmetric point $\phi=0$ is questionable independent of our
approximation: Since the  negative mass-squared is greater than the
cutoff-squared, the symmetric phase  cannot exist in any physical sense. In
fact in this regime \cactusflow\ is not the limit of the equations with smooth
cutoff because the continuity condition in lemma \lemma\ is violated for some
$q<\Lambda_0$. If one figures out from \smoothflow\ the smooth cutoff form of
the cactus approximation then the R.H.S. is divergent if $\sigma(;t)<-1$, for
all $\epsilon>0$.

The next step in the approximation method is to expand in momentum scale. We
see from \twoflow\ that  the $n=4$ flow equation  is needed to determine the
coefficients. From \scaledflow\ we obtain
\eqn\fourflow{
\eqalign{
&\left( {\partial\over\partial t}+p_i^\mu{\partial\over\partial p_i^\mu}\right)
\gamma(\p_1,\cdots ,\p_4;t)=
{1\over(4\pi)^2} {1\over1+\sigma(1;t)} <\gamma(\q,-\q,\p_1,\cdots,\p_4;t)
>_{q=1}\cr
&-{2\over(4\pi)^2}{1\over1+\sigma(1;t)}<
G(r_{12};t)
\gamma(\q,-\r_{12},\p_1,\p_2;t)\gamma(\r_{12},-\q,\p_3,\p_4;t)
\cr
&\phantom{-{2\over(4\pi)^2}{1\over1+\sigma(1;t)}<
G(r_{12};t)
\gamma(\q,-\r_{12},\p_1,\p_2;t)\gamma(\r_12,}
+(2\leftrightarrow3)+(2\leftrightarrow4)>_{q=1}\cr}}
where $\r_{ij}=\q+\p_i+\p_j$, and in the second average the last two terms  are
the same as the first term but with indices swopped as indicated.
To truncate at this level we set $\gamma(\p_1,\cdots,\p_6;t)\equiv0$. This is a
sensible approximation because, even if the 6-point vertex is non-zero in the
bare action,  it is irrelevant at the gaussian fixed point and therefore at
sufficiently low energy-scales it shrinks rapidly as $\Lambda$ falls, until it
is determined to good approximation as a perturbative series in the now small
renormalized coupling $\lambda$. Note that since the R.H.S. of \fourflow\ is
positive in this truncation we can see already that the 4-point vertex must
shrink as we flow to low energies.

 We now make the theory massless by setting $\sigma(p;t)\equiv0$. This is not
meant to be taken seriously as an approximation: we do so purely because it
greatly simplifies the equations since we can now study \fourflow\ without
having to worry explicitly about \twoflow. With these changes we  have
\eqn\truncflow{
\eqalign{
&\left( {\partial\over\partial t}+p_i^\mu{\partial\over\partial p_i^\mu}\right)
\gamma(\p_1,\cdots ,\p_4;t)=\cr
&-{2\over(4\pi)^2}<
{\theta(r_{12}-1)\over r_{12}^2}
\gamma(\q,-\r_{12},\p_1,\p_2;t)\gamma(\r_{12},-\q,\p_3,\p_4;t)\cr
&\phantom{-{2\over(4\pi)^2}<
{\theta(r_{12}-1)\over r_{12}^2}\gamma(\q,-\r_{12},\p_1,\p_2;t)\gamma\r_12}
 +(2\leftrightarrow3)+(2\leftrightarrow4)>_{q=1}\quad.\cr}}
Determining the perturbative solution,  we will find that this truncation is
exact at one-loop. Again the perturbative expansion is easiest performed by
integrating both
sides of \flowexpanded\ with respect to $\Lambda$ using the b.c.
$\Gamma(\p_1,\cdots,\p_4;\Lambda_0)=\lambda_0$, and  iterating. The Feynman
graphs up to two-loops are given in fig.\fig\twoloop{ }.
\midinsert
\centerline{
\psfig{figure=exactrgfig5.ps,width=6in}}
\bigskip
\vskip 0.3cm
\centerline{\vbox{{\bf Fig.5.} Feynman graphs
contributing to the 4-point vertex, in the truncation defined by setting the
6-point vertex to zero (with the self-energy also set to zero).}}
\endinsert
The one-loop contribution to $\Gamma(\p_1,\cdots,\p_4;\Lambda)$  is found to be
$$-\lambda_0^2\sum_{i=2}^4\int\!\!{d^4q\over(2\pi)^4}\
{[\theta(q-\Lambda)-\theta(q-\Lambda_0)]\theta(|\q+\p_1+\p_i|-q)\over
q^2(\q+\p_1+\p_i)^2}\quad.$$
 Substitute $\q\to\q-\p$ followed by $\q\to-\q$ and add the result to the above
(dividing the whole by 2). After a little rearrangement this is seen to be
$$-{\lambda_0^2\over2}\sum_{i=2}^4\int\!\!{d^4q\over(2\pi)^4}\
{\theta(q-\Lambda)\theta(|\q+\p_1+\p_i|-\Lambda)-\theta(q-\Lambda_0)\theta(|\q+
\p_1+\p_i|-\Lambda_0)\over
q^2(\q+\p_1+\p_i)^2}\quad.$$
This is the exact one-loop answer, with however a U.V. cutoff that only works
when {\sl all} momenta in the loop are larger than $\Lambda_0$. This is a
reflection of our dropping the explicit overall cutoff at the end of sect.3.
Similarly the first two-loop diagram in fig.\twoloop\ is given exactly (as
follows from adding several contributions after exchanging orders of
integration), while some of the second is missing because it is  provided by
the one-loop 6-point vertex (as indicated by the dotted box).

Since eqn.\truncflow\ incorporates the exact one-loop result it gives in
particular the right one-loop divergence $-{3\lambda_0^2 t/ (4\pi)^2}$. (Recall
that $t\equiv\ln(\Lambda_0/\Lambda)$). If the truncation is indeed
perturbatively renormalizable (\cf sect.2) then  the leading log divergences
$\sim\lambda_0^m t^{m-1}$, for all $m$, must be those determined by the
one-loop renormalization group. Clearly these leading log terms come from
iterating the constant part of $\Gamma(\p_1,\cdots,\p_4;\Lambda)$, \ie dropping
all external-momentum dependent pieces. But substituting a constant
$\Gamma(\p_1,\cdots,\p_4;\Lambda)=\lambda(t)$ in such a way into \truncflow\
gives:
\eqn\zeroth{
{d\lambda(t)\over d t}=-{3\over(4\pi)^2}\lambda^2(t)\quad,
}
\ie  the one-loop $\beta$-function as required.

It is an instructive exercise  to compute the two-loop divergences. This will
provide further explicit checks on the method. Two-loop diagrams of the first
type in fig.\twoloop\ are straightforward and give in total
$3\lambda_0^3t^2/(4\pi)^4$. The last type of two-loop diagram gives in total
$$6\lambda_0^3\int^{\Lambda_0}_{\Lambda}\!\!{d^4p\over(2\pi)^4}\
{1\over p^4}\int^{\Lambda_0}_{p}\!\!{d^4q\over(2\pi)^4}\
{\theta(|\q+\p|-q)\over q^2 (\q+\p)^2}\quad.$$
(We have set $\p_1,\cdots,\p_4=0$ since they do not appear in the divergences).
The inner integral is not doable exactly. We follow the momentum scale
expansion \expgam--\average\ which here means expanding  the inner  integrand
in $p/q$ and performing the angular integration. One finds
\eqn\momexp{
\int^{\Lambda_0}_{p}\!\!{d^4q\over(2\pi)^4}\
{\theta(|\q+\p|-q)\over q^2 (\q+\p)^2}=
{1\over4\pi^3}\int^{\Lambda_0}_{p}\!\!{dq\over q}\left\{{\pi\over4}
-{1\over6}{p\over q}
-{1\over240}\left({p\over q}\right)^3-{3\over8960}\left({p\over q}\right)^5
-\cdots\right\}
}
which, apart from the constant term, is an expansion in odd powers of $p$ (as
follows from taking the symmetric part under $\p\to-\p$ and using
$<{1\over(\p+\q)^2}>_{\q=q>p}={1\over q^2}$). Integrating, and substituting
into the outer loop integral one obtains the $O(\lambda_0^3)$ contribution
to $\Gamma(\p_1,\cdots,\p_4;\Lambda)$ as
$${9\lambda_0^3\over(4\pi)^4}\left(\ln{\Lambda_0\over\Lambda}\right)^2-\eta
{\lambda_0^3\over (4\pi)^4}\ln{\Lambda_0\over\Lambda}+\ finite$$
where $\eta$ is given as a rapidly convergent numerical series
$\eta={1\over\pi}(8+{1\over15}+{9\over2800}+\cdots)$. The partial sums
$\eta=2.5465,2.5677,2.5687,\cdots,2.568818$ converge to 3sf already at order
$p^3$ in the momentum expansion, after which approximately an extra decimal
place in accuracy is added with each new term. (The ``$finite$'' term converges
somewhat faster).  {\sl Nota bene} that, while the terms bounded by curly
brackets in \momexp\ may be regarded as an expansion in the small quantity
$p/q$, once the integral is performed there are no longer identifiable small
parameters to control the approximation (because the lower limit is $p$).
Nevertheless, as we have seen, the expansion converges very rapidly. The
leading log divergence is as expected, while the subleading divergence implies
the $O(\lambda^3)$ term of the $\beta$-function:
\eqn\betatwo{
{d\lambda(t)\over dt}=-{3\over(4\pi)^2}\lambda^2(t)
+{\eta\over(4\pi)^4}\lambda^3(t)+O(\lambda^5)\quad.
}

Now we compute the momentum scale expansion \nonp ly on \truncflow. The
$O(\rho^0)$ case is  evidently solved by
$\gamma_0(\p_1,\cdots,\p_4;t)=\lambda(t)$, \ie momentum independent, and gives
the one-loop $\beta$-function \zeroth\ as the flow equation. Following the
prescription \expgam--\average\ we see that the $O(\rho^1)$ case generates,
from the expansion of the propagator and I.R. cutoff, the non-analytic terms
$|\p_1+\p_2|$, $|\p_1+\p_3|$ and $|\p_1+\p_4|$
\ie $\sqrt{S}$, $\sqrt{T}$ and $\sqrt{U}$ in Mandelstam variables. Therefore we
try
\eqn\ouransatz{
\gamma(\p_1,\cdots,\p_4;t)=\lambda(t)
+\gamma_1(t)\,(\sqrt{S}+\sqrt{T}+\sqrt{U})\quad.}
 This gives, for example, (working always to order $\rho^1$):
$$\gamma(\q,-\q-\p_1-\p_2,\p_1,\p_2;t)\equiv\lambda
+2\gamma_1+\sqrt{S}\,\gamma_1(1+x)\quad,$$
where $x=(\p_1+\p_2).\q/\sqrt{S}$. Substituting all expansions into \truncflow\
and evaluating the average we find
that \ouransatz\ is indeed the solution providing $\lambda(t)$ and
$\gamma_1(t)$ satisfy:
\eqna\first
$$\eqalignno{
{d\lambda\over dt} &= -{3\over(4\pi)^2}(\lambda+2\gamma_1)^2 &\first a\cr
{d\gamma_1\over dt}&=
-\gamma_1+{1\over24\pi^3}(\lambda+2\gamma_1)(\lambda-[3\pi+2]\gamma_1)\quad,
&\first b\cr
}$$
with boundary conditions $\lambda(0)=\lambda_0$ and $\gamma_1(0)=0$.

Integrating these equations numerically one finds that they focus in on the
gaussian fixed point
($\lambda=\gamma_1=0$) as required. This is illustrated in
fig.\fig\focus{ }.
\midinsert
\centerline{
\psfig{figure=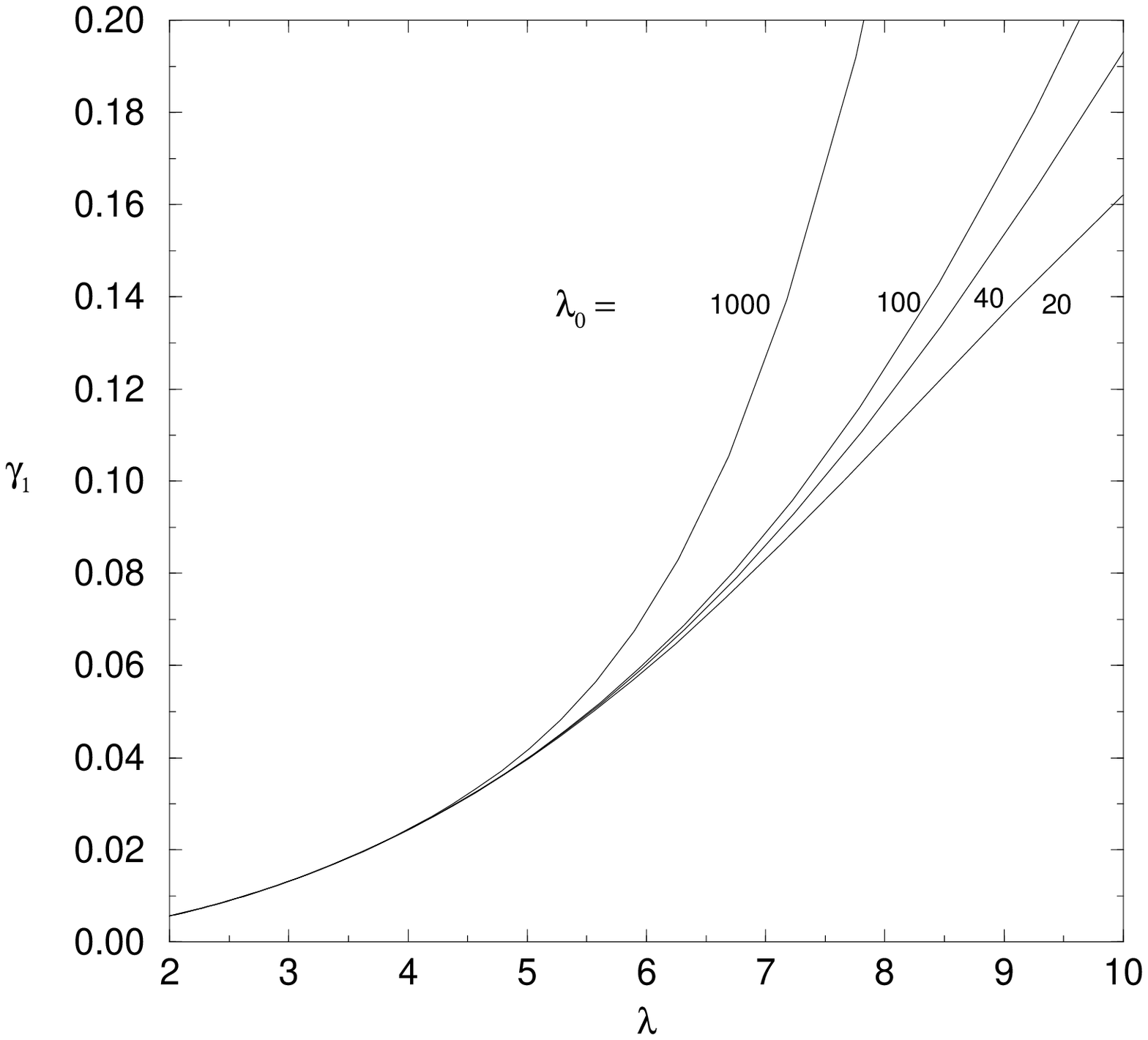,width=6in}}
\vskip -2.3in
\centerline{\vbox{{\bf Fig.6.} $\gamma_1$ plotted as a function of $\lambda(t)$
for  bare
couplings $\lambda_0=20,40,100,1000$.
}}
\endinsert
For small enough $\lambda$ the
solution for $\gamma_1$ is well approximated by the renormalization group
improved perturbation expansion for this truncation, which is readily derived
by using \first{a}\ to solve  \first{b}\ for $\gamma_1$ as a power series in
$\lambda(t)$. One finds
\eqn\RGpert{\gamma_1={1\over24\pi^3}\lambda^2+{1\over96\pi^5}\lambda^3
+\left(
{ {7}\over{1536\,\pi ^{7}}}+{ {5}\over{2304\,\pi ^{8}}}-{{1}\over
{3456\,\pi ^{9}}} \right)\lambda^4+\cdots}
which substituted into \first{a}\ gives the $\beta$-function for this
truncation:
\eqn\betafunc{
{d\lambda\over dt}=-{3\lambda^2\over(4\pi)^2}
-{8\over\pi}{\lambda^3\over(4\pi)^4}-\left({32\over\pi}
+{16\over3\pi^2}\right){\lambda^4\over(4\pi)^6}-\cdots\quad.}
Several things should be noted about this solution. Firstly it is universal
(\ie independent of the cutoff). It may be shown that the \nonp\ corrections
from \first{}\ are all cutoff dependent. Secondly \RGpert\ and \betafunc\ are
infinite series in the renormalised coupling constant. Therefore they
incorporate non-trivial contributions to all loops. Thirdly, by comparison with
\betatwo\ one sees that already at  first  order in the momentum scale
expansion 99.1\% of the two-loop contribution is included. In fact it may be
shown that these series are asymptotic of the expected form, so they already
capture much of the qualitative behaviour of perturbation theory to all orders.

Now we briefly investigate the \nonp\ behaviour to test the convergence of the
momentum scale expansion in this regime. Plotted in fig.\fig\zf{ }\ is a
comparison  between the
zeroth order \zeroth\ and first order \first{}\ results for $\lambda(t)$, given
the extreme bare coupling $\lambda_0=1000$.
\midinsert
\centerline{
\psfig{figure=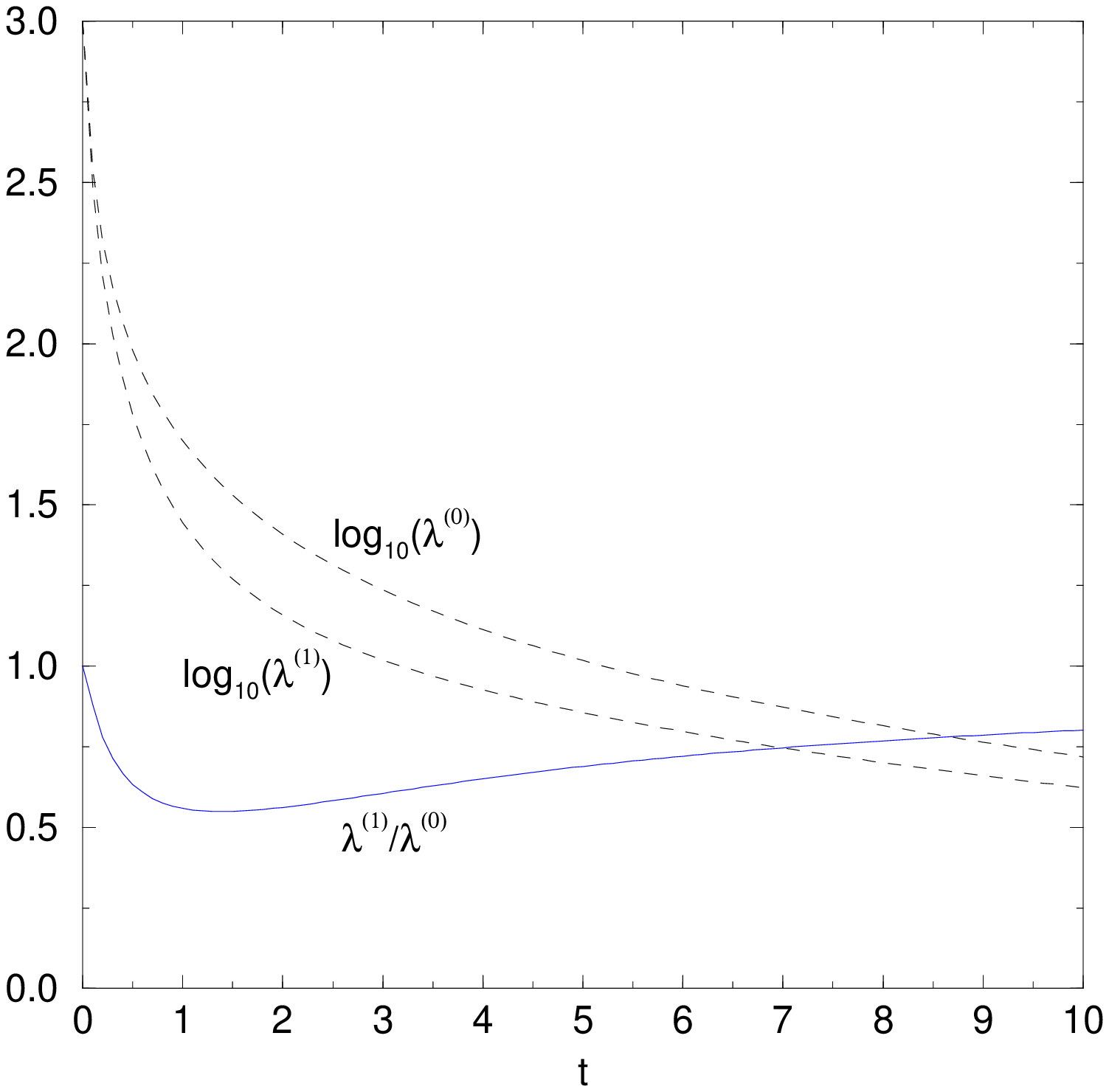,width=6in}}
\vskip -2.3in
\centerline{\vbox{{\bf Fig.7.} A comparison
between zeroth $\lambda\equiv\lambda^{(0)}(t)$ and first order
$\lambda\equiv\lambda^{(1)}(t)$ approximations to $\lambda(t)$ in the momentum
scale expansion, for bare coupling $\lambda(0)=\lambda_0=1000$. Plotted are
$\log_{10}\lambda^{(0)}$, $\log_{10}\lambda^{(1)}$ and
$\lambda^{(1)}/\lambda^{(0)}$ against $t$.
}}
\endinsert
We see that the difference is never
large and falls to zero (as it must) as $t\to\infty$.
Since we are never interested in scales where cutoff effects dominate we
quantify the difference in, what L\"uscher and Weisz refer to as, the ``scaling
region'' where the computations are universal to a good approximation. For us
this corresponds to the region $\lambda(t)<\lambda_s$ where, say,
$\lambda_s\sim5$
for $\lambda_0\le1000$, since cutoff effects are then $\le 5\%$. (They reach
$\approx$5\% at  $\lambda_0=1000$ for $\lambda=5$, and are computed by
comparing $\gamma_1(t)$ with \RGpert). In this region all quantities may be
expressed in terms of $\lambda(t)$, whose evolution is given by \betafunc:
$$t=t_\infty
+{16\pi^2\over3\lambda(t)}+{8\over9\pi}\ln\lambda(t)+O(\lambda)\quad.$$
The integration constant $t_\infty$, the only remaining \nonp\ quantity, is
readily computed numerically using the above formula. We find to first order in
the momentum scale expansion
$t_\infty=-2.98$ for $\lambda_0=1000$, while $t_\infty=-3.48$ for
$\lambda_0=20$, for example. These should be compared to the (zeroth order)
one-loop result
$t_\infty=-{16\pi^2/(3\lambda_0)}$. The effect on low energy is given by
$\Delta\lambda/\lambda\approx (16\pi^2/3)\lambda\,\Delta t_\infty$. Expressed
as a percentage this is $\Delta\lambda/\lambda=a\lambda$ where $a=-5.6\%$,
$-1.7\%$, for $\lambda_0=$ 1000, 20, respectively.

Finally we briefly consider the more realistic case where $\sigma(p;t)$ is not
set to zero, and \fourflow\ is therefore coupled to \twoflow. From \ouransatz\
one might expect a momentum scale expansion of the form
$\sigma(p;t)=\sigma_0(t)+p\sigma_1(t)+O(p^2)$. If so, by dimensions one would
expect $\sigma_1$ to be linearly divergent \ie $\sigma_1\sim
\Lambda_0/\Lambda$. This would be a disaster, as already explained at the
beginning of sect.4. In fact $\sigma_1$ is identically zero, since substituting
\ouransatz\ in \twoflow\ gives momentum dependence of the form
$<|\q+\p|+|\q-\p|>_{q=1}$, yielding  an expansion in $p^2$.

\section{Summary.}
The main points of the paper are briefly recapitulated.

The aim was to find a method of continuum calculation in realistic quantum
field theories consisting of a sequence of better and better approximations,
calculable without inhuman effort, and applicable even if there are no
obviously
identifiable small parameters to control the approximation.

The Wilson renormalization group framework was chosen primarily because, unlike
other frameworks,  truncations are guaranteed to be at least perturbatively
renormalizable.

Stated completely, the most promising method may be summarised as follows:
Use the differential flow equation in $\Lambda$ for the Legendre effective
action\foot{Here and later the effective actions are defined minus the
classical kinetic term which incorporates the cutoff.}\ $\Gamma_\Lambda[\phic]$
defined by a theory with I.R. momentum cutoff $\Lambda$, and derived (formally)
by ignoring the overall U.V. cutoff.  Take the sharp cutoff limit in which the
I.R. momentum cutoff is just a Heaviside $\theta$-function.
$\Gamma_\Lambda[\phic]$ may be thought of as equivalent to a Wilsonian
effective action with effective U.V. cutoff $\Lambda$. The boundary condition
to the flow equation is supplied at $\Lambda=\Lambda_0$ by a
$\Gamma_{\Lambda_0}[\phic]$, chosen to be local, which may be thought of as
equivalent to a bare action.
Finally, approximate the flow equation by truncation  at some $n$-point 1PI
(one-particle irreducible) greens function and  expansion of each $m$-point
greens function in momentum scale up to some maximum power.

This prescription was arrived at as follows. The Wilson effective action
$S_\Lambda[\phi]$, where $\Lambda$ is an effective U.V. cutoff, was shown to
have a reinterpretation in terms of the generator of connected greens functions
with I.R. cutoff $\Lambda$. As such it has an expansion in terms of 1PI
greens functions, and thus is closely related to the Legendre effective action
$\Gamma_\Lambda[\phi]$ in a theory with I.R. cutoff $\Lambda$. Unlike
$S_\Lambda[\phi]$, $\Gamma_\Lambda[\phi]$ is insensitive to the precise form of
the cutoff when the width $\epsilon$ of the cutoff is small. It does appear to
depend on an overall momentum cutoff $\Lambda_0$ however, but we show that a
tree-level reparametrization invariance allows us to change $\Lambda_0$. We use
this invariance to interpret the flow equation for $\Gamma_\Lambda[\phic]$ in
the simplifying limit $\Lambda_0\to\infty$. This interpretation consists of
defining the theory by a local `bare' $\Gamma_{\Lambda_0}[\phic]$, the initial
condition for the flow equation (from which a bare action and finite overall
momentum cutoff can be reconstructed if desired).

The 1PI greens functions, in the sharp cutoff limit  $\epsilon=0$, are
non-analytic in the momenta around $\p={\bf0}$ corresponding to non-local
behaviour, however this appears merely to be a technical problem and not one of
 principle.  On the other hand the smooth cutoff equations do not appear to
have any useful approximation, the local derivative expansion having severe
problems with convergence and  sensitivity to the form of the cutoff.
Straightforward
numerical integration, expansion in $\ln(\Lambda_0/\Lambda)$, and expansion in
$\Lambda$, are considered but are probably of limited use.

Two simple model examples  were considered to test the most promising method:
the ladder (or rather cactus) approximation, and
 a model incorporating the first irrelevant correction to the
 renormalized coupling, both for four dimensional \pf. The cactus approximation
is the simplest possible non-trivial \nonp\ approximation. It gives sensible
qualitative results and even
 yields a maximum renormalized
coupling in excellent quantitative agreement with previous precision work.
For a certain range of bare couplings in the unstable symmetric phase it gives
peculiar results, however it is easily seen that the approximation method does
not converge there. The problem is avoided by shifting to the symmetric phase,
or it may be resolved in other ways.
In the truncation defined by setting the 6-point vertex to zero, and further
simplified by setting the self-energy to zero, the two-loop contribution was
computed at zero external momentum using the momentum scale expansion, and seen
to converge very rapidly, despite the fact that it is a numerical series with
no obviously identifiable
expansion parameter. Non-perturbatively the zeroth order in the momentum scale
expansion coincides with the one-loop $\beta$-function.   First  order in the
momentum expansion gives a pair of flow equations whose (renormalization group
improved)
perturbation expansion yields an asymptotic power series
for the $\beta$-function, incorporating
 99.1\% of the expected two-loop contribution.  The \nonp\ corrections compared
to the zeroth order  result are small,  giving confidence in a fuller use of
these
approximation methods for obtaining accurate results.

A full demonstration that the proposed approximation method meets our
requirements of convergence and calculability awaits a proper calculation of
some real \nonp\ problem. Research on using this to compute triviality bounds
for the Higgs mass is underway. Since the method is tied to momentum cutoff,
further work is needed before it can be applied to gauge theories.

\bigskip\bigskip

\centerline{\bf Note Added.}
Since this paper was submitted, a number of other relevant works have come to
our attention. We here give a more complete comparison with
earlier work.
Of course there has been a vast amount of work on the exact
renormalisation group in various guises; see for example ref.\kogwil\ for
an early review and  references. Continuum flow equations with smooth cutoff
were derived in ref.\kogwil\ and are surely
 equivalent to those of Polchinski\pol\ (as indeed stated in ref.\pol). The
sharp
cutoff equations were derived by Wegner and
Houghton\ref\wegner{F.J. Wegner and A. Houghton, Phys. Rev. A8 (1973) 401.},
although the ambiguities we discussed in this paper were avoided by
formulating the equations only for discrete momenta. It is this,
presumably, that discouraged a more wide-spread use of these equations.
The formal continuum limit for essentially these equations was derived in
 the appendix of ref.\ref\weinberg{ ``Critical Phenomena
for Field Theorists'', S. Weinberg,
lectures, Erice Subnucl. Phys. (1976) 1.}.
Weinberg noted that the equations were unpleasant, and by re-expressing the
vertices as a tree expansion in some new vertices,
derived  flow equations for the new vertices in the sharp cutoff limit.
The transformation in terms of trees was used in ref.\ref\morgan{D. Morgan,
Ph.D.
thesis (1991) University of Texas, Austin.}\ to establish equivalence between
these equations and Polchinski's equations directly, for smooth cutoff also.
Weinberg's equations coincide with the flow equations of the 1PI greens
functions given in this
paper (and refs.\bonini\wet), only the interpretation (in terms of
greens functions as given here) is missing. Flow equations for the Legendre
effective action (as given here and in refs.\bonini\wet) were derived in
ref.\ref\nicii{J.F. Nicoll and T.S. Chang,
Phys. Lett. 62A (1977) 287.}. I find it surprising that these equations appear
not
to have been used for approximations other than rederiving the $\epsilon$
expansion\wegner\ and constructing the simplest ``$p^0$'' approximation where
all momentum dependence is discarded\wegner\jf-\el. This latter equation was
probably
first explicitly written down in ref.\ref\niciii{J.F. Nicoll, T.S. Chang and
H.E.
Stanley, Phys. Lett. 57A (1976) 7.}, the leading order large $N$ case is
 given in ref.\wegner, see
ref.\ref\KIG{K. Kawasaki, T. Imaeda, and J. Gunton, in
``Perspectives in Statistical Physics'', ed. H. Ravech\'e, North-Holland
(1981).}\
for a review.
In this respect mention should
also be made of Wilson's approximate recursion formula\kogwil\ which, like
the $p^0$ approximation, also drops all momentum dependence, but here in an
equation for
halving the cutoff. As far as I know it gives similar results to $p^0$
approximation\has\kogwil, but in contrast  to the $p^0$ approximation
 it cannot be regarded as the first term in a sequence of successive
approximations\kogwil . It also differs in that the authors of ref.\kogwil\
prefer to
consider smooth rather than sharp cutoffs. The reasons given are the
difficulties inherent in using the Wegner-Houghton equation\wegner\
and an understandable ``philosophical'' prejudice against the induced
non-localities. As stated earlier in the paper, the transformation
to 1PI vertices resolves the former's difficulties while, despite searching,
we have found no deep problems with the latter. Also it should again be noted
that differences between
a sharp and smooth cutoff are in any case of no relevance to the
 $p^0$ approximation.

\acknowledgements

It is a pleasure to thank the following people for their interest and
discussions: Michel Bauer, Patrick Dorey,  Simon Hands, Tim Hollowood, Luis
Miramontes and  Moshe Moshe.

%\listfigs
\listrefs
\end